\documentclass[11pt]{article}

\usepackage[preprint]{acl}  
\usepackage{times}
\usepackage{latexsym}
\usepackage[T1]{fontenc}
\usepackage[utf8]{inputenc}
\usepackage{microtype}
\usepackage{inconsolata}
\usepackage{graphicx}
\usepackage{booktabs}
\usepackage{multirow}
\usepackage{amsmath}
\usepackage{amssymb}
\usepackage{enumitem}
\usepackage{xcolor}
\usepackage{fancyvrb}
\usepackage{array}
\usepackage{algorithm}
\makeatletter  
\renewcommand\fs@ruled{\def\@fs@cfont{\bfseries}\let\@fs@capt\floatc@ruled
  \def\@fs@pre{\hrule height.8pt depth0pt \kern1.2pt \hrule height.4pt depth0pt \kern2pt}%
  \def\@fs@post{\kern2pt\hrule height.4pt depth0pt \kern1.2pt \hrule height.8pt depth0pt \relax}%
  \def\@fs@mid{\kern2pt\hrule\kern2pt}%
  \let\@fs@iftopcapt\iftrue}
\makeatother
\restylefloat{algorithm}
\usepackage{algpseudocode}
\newcommand{\figsafe}[2]{\IfFileExists{#2}{\includegraphics[#1]{#2}}{\IfFileExists{figures/#2}{\includegraphics[#1]{figures/#2}}{\IfFileExists{generated_nb/#2}{\includegraphics[#1]{generated_nb/#2}}{\IfFileExists{../generated_nb/#2}{\includegraphics[#1]{../generated_nb/#2}}{\fbox{\parbox{0.9\linewidth}{\centering\ttfamily missing figure: \detokenize{#2}}}}}}}}

\algrenewcommand{\algorithmicindent}{1.2em}
\algrenewcommand{\algorithmiccomment}[1]{\hfill{\footnotesize\color{gray}\textit{#1}}}

\newcommand{\ours}{CQ2-CARE}

\newcommand{\rev}[1]{#1}  

\title{Skip the Compressor:\\Coding Agents Fix More Issues When They Pick Whole Files}

\author{Xiaocong Yang$^*$ \quad Rishit Chatterjee$^*$ \quad Himanshu Udupi \\
  \textbf{Vikram Ramavarapu} \quad \textbf{ChengXiang Zhai} \\
  University of Illinois Urbana-Champaign \\
  \texttt{\{xy51, rc33, hudupi2, vikramr2, czhai\}@illinois.edu}}

\begin{document}
\maketitle
{\renewcommand{\thefootnote}{*}\footnotetext{Equal contribution.}}

\begin{abstract}
A coding agent works on repositories that can be far larger than its context window can hold.
Prompt compressors and retrievers offer solutions to this problem, but patching an issue needs two things at once: the files that must change, and their code exactly as written, since the model must reproduce the lines it edits.
Retrievers rank code by its similarity to the issue rather than by whether it must change, and compressors delete text to fit the budget.
We show that at an equal token budget, an agent resolves more issues when the budget goes to choosing which files to show verbatim, as our simple selector \ours{} does, than when it goes to pruning tokens or retrieving chunks.
We fit every context to the same per-issue budget on 300 recent SWE-rebench issues and grade each patch with the benchmark's tests.
With GPT-5.6 and Claude Opus 5, \ours{} resolves 43.7\% and 46.7\% of the issues, 6.3 points more than the best of three compressors with either model, and 9.3 and 12.7 points more than the best of three retrievers.
\rev{\ours{} even resolves more issues than two methods trained for issue resolution, SweRank (by 0.7 and 1.7 points) and SWE-Pruner (by 2.0 and 7.3).}
Our analysis shows the defects of the baseline methods: retrievers often miss the files to edit, and compressors keep them but alter their text, so fewer patches apply. Exact-file selection needs no compression model and shows every line as the repository has it, so the model can quote the code it edits.
\end{abstract}

\section{Introduction}
\label{sec:intro}

Agents based on Large Language Models (LLMs) now resolve real software issues end to end: they read an issue, explore a repository and write a patch that the project's tests accept \citep{jimenez2024swebench,yang2024sweagent,wang2025openhands,xia2025agentless}.
Managing what enters their context already makes them cheaper: it can halve an agent's cost \citep{lindenbauer2025complexity} or cut it by a fifth to a third \citep{xiao2026agentdiet}.
Every prompt token is paid for on every call, and models use information in the middle of a long input less reliably than at its ends \citep{liu2024lostinthemiddle}.

\begin{figure*}[t]
\centering
\figsafe{width=\textwidth}{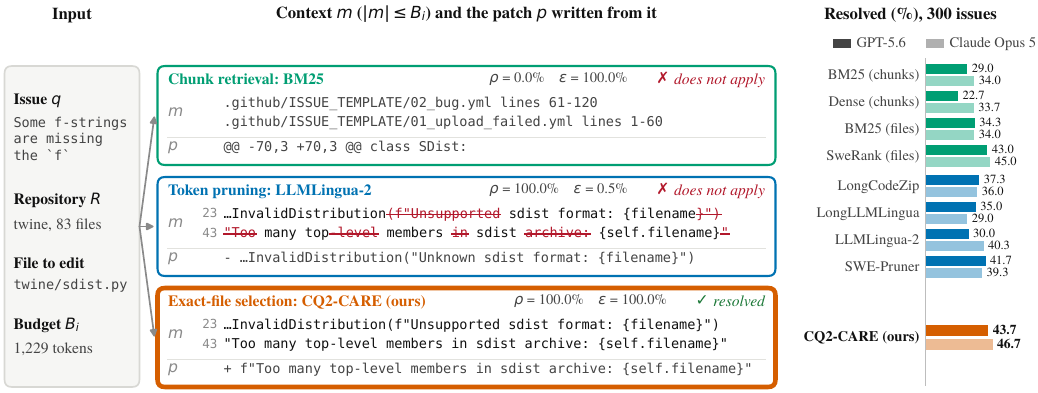}
\caption{\textbf{An issue sample under different compression methods.} The issue (\texttt{pypa\_\_twine-1225}) reports error messages that lack the \texttt{f} of an f-string; the gold patch edits \texttt{twine/sdist.py}. $\rho$: gold-file recall, the share of the gold-patch files in the context $m$; $\varepsilon$: the share of $m$'s lines that are verbatim. BM25 fills $B_i$ with issue templates (two of three spans shown), so its patch guesses lines it never saw; LLMLingua-2 deletes tokens (struck out), among them the \texttt{f} of line 23, so its patch rewrites a line that does not exist; ours shows the file verbatim. Indentation is trimmed; an ellipsis marks an elided prefix. Chosen for illustration as the smallest of the 4 such issues out of 300 (a random one is in Appendix~\ref{app:example}). Right: resolution with GPT-5.6 (solid) and Claude Opus 5 (light), every context fitted to $B_i$\rev{. SWE-Pruner, a line pruner trained for coding agents, is the last bar of the pruning group, and SweRank, a trained retriever of whole functions, the last bar of the retrieval group}.}
\label{fig:main}
\end{figure*}

Yet every step still needs code in the prompt, and a coding agent's context window is small next to the repositories it works on: SWE-bench codebases average 3,010 non-test files and 438K lines \citep{jimenez2024swebench}.
We therefore treat the repository as the agent's memory and the prompt as a compressed view of it, one that leaves most of the code out.
To resolve an issue, the view still has to keep two things: the files the fix must edit, and their text exactly as it is.
For anyone building a coding agent, this raises a practical question: \emph{at a fixed token budget, which compression of the repository lets the model resolve the most issues?}

Three lines of work each study part of this question: prompt compression for long natural-language prompts \citep{jiang2023llmlingua,pan2024llmlingua2}, retrieval of repository code for code completion and issue resolution \citep{zhang2023repocoder,jimenez2024swebench}, and issue-resolution systems that prune what an agent reads \citep{wang2026swepruner} or compress the context around known fault locations \citep{jia2026swezze}.
Figure~\ref{fig:main} shows how the first two can fail on one issue from our study, a report that some error messages lack the \texttt{f} of an f-string.
A chunk retriever fills the budget with the repository's issue templates, which share the report's wording, and never shows the file to edit.
A token pruner keeps that file but deletes \texttt{f} prefixes and quotation marks, so the model's patch rewrites a line that does not exist and fails to apply.
Shown the file whole and verbatim, the same model writes the fix.
None of these lines of work compares retrieval, token pruning and file selection at a matched budget per issue, with each method free to decide which files enter the context.

Both failures follow from how these tools work.
Retrievers rank chunks or files by their similarity to the issue \citep{robertson2009bm25,jimenez2024swebench,suresh2025cornstack}, so text that echoes the report can outrank the file to edit.
Prompt compressors delete the tokens or code blocks that a small model scores as least useful \citep{jiang2023llmlingua,jiang2024longllmlingua,pan2024llmlingua2,shi2025longcodezip}; this costs little when the output is an answer or a summary, but an edit must quote the code it changes exactly \citep{xia2025agentless,anthropic2024swebench,aider-repo}.

The remedy follows: spend the budget on choosing which files to show, and show them whole and verbatim, so that edits quote text that exists and no compression model runs in the loop.
Since each chosen file is shown whole, the selector must name few files, and the right ones.
We therefore build a simple selector, \ours{}, that asks the model under test twice, independently, which files must change and keeps the files both answers name, taking agreement between samples as a sign of reliability \citep{wang2023selfconsistency}; it then adds the short spans that grep finds for definitions, callers or strings the model requests, so code the fix depends on is shown verbatim even outside the chosen files.
Its strength is its choice of unit, not its sampling scheme: a single answer cut to the same number of files does about as well, so what matters is choosing few files well (\S\ref{sec:selector}).

We test this remedy in a controlled comparison on 300 issues from the SWE-rebench leaderboard \citep{badertdinov2025swerebench}, one per repository: a fixed model receives the issue and one context fitted to the size of the \ours{} context for that issue, writes one patch, and the benchmark's harness grades it with the issue's tests.
We describe each context by its gold-file recall, the share of the files to edit that it contains, and by the share of its lines that are verbatim, and we run controlled tests that change one of the two at a time.

With each of two model families, the selector resolves more issues than each of three compressors and three retrievers, among them SWE-bench's file-level BM25; with Claude, its lead over one compressor, LLMLingua-2, is narrow.
The losses separate cleanly: retrievers keep the text verbatim but often miss the files to edit, and token pruners keep those files but alter their text, so fewer patches apply.
Controlled tests show that each loss costs on its own: deleting tokens from the selector's own contexts, with the files unchanged, costs about as much as the compressors lose in the main comparison, while a compressor that keeps whole lines matches the selector when it starts from the same files.

\rev{The selector even resolves more issues than two methods trained for issue resolution: SWE-Pruner, a line pruner, and SweRank, a retriever that shows whole functions verbatim and comes within two points.}

\noindent\textbf{Our contributions are as follows:}
\begin{itemize}[leftmargin=*,itemsep=1pt,topsep=2pt]
\item \ours{}, a simple selector that spends the context budget on the files the model names and shows them verbatim, with no compression model (\S\ref{sec:setup}).
\item A budget-matched, test-graded comparison of \ours{} with three prompt compressors, three retrievers, among them SWE-bench's file-level BM25, \rev{and two methods trained for issue resolution,} on 300 issues and two model families (\S\ref{sec:main}).
\item A decomposition of the losses into gold-file recall and verbatim lines, with controlled tests that change one at a time (\S\ref{sec:lose}--\ref{sec:controlled}).
\end{itemize}

\section{Context as Compressed Memory}
\label{sec:setup}

Let $R$ be the repository at the issue's base commit (the agent's memory), $q$ the issue text and $B$ a token budget.
A context step $C$ maps them to a context $m = C(q, R) = (s_1, \dots, s_K)$ of $|m| \le B$ tokens, a sequence of source spans.
Span $s_k$ shows lines $a_k$ to $b_k$ of a file $f_k$ of $R$ under a header \texttt{-{}-{}- SOURCE k: path lines a-b -{}-{}-}; we write $f[a{:}b]$ for lines $a$ to $b$ of $f$ verbatim, so a whole file is $f[1{:}|f|]$.
A fixed model $\pi$ then writes a patch $p = \pi(q, m)$, and $y \in \{0,1\}$ records whether the issue is resolved, i.e.\ whether every FAIL\_TO\_PASS and PASS\_TO\_PASS test passes after applying $p$; we compare context steps by their resolution rate, the mean of $y$ over issues, at equal budgets.

The three rows of Figure~\ref{fig:main} are the three ways to fill the budget.
\emph{Chunk retrieval} splits $R$ into units (top-level definitions, line windows or whole files), ranks them against $q$, with $u_{(j)}$ the $j$-th best, and packs them verbatim in rank order: the top $r$ whole units that fit in $B$, then $\bar u_{(r+1)}$, the longest whole-line prefix of the next unit that still fits:
\begin{equation}
  m_{\mathrm{ret}} = \big(u_{(1)}, \dots, u_{(r)}, \bar u_{(r+1)}\big).
  \label{eq:ret}
\end{equation}
\emph{Token pruning} shortens a larger candidate context $\tilde m$ by deleting the units (tokens, lines or code blocks) that a scoring model rates least useful; with $D_r(\tilde m)$ the units that the method's own rule deletes at rate setting $r$, and $r$ chosen so that $|m| \approx B$:
\begin{equation}
  m_{\mathrm{prune}} = \tilde m \setminus D_r(\tilde m).
  \label{eq:prune}
\end{equation}
\begin{table*}[t]
\centering
\footnotesize
\setlength{\tabcolsep}{3.2pt}
\begin{tabular*}{\textwidth}{@{\extracolsep{\fill}}lrrrrrr@{\extracolsep{0pt}}l@{\extracolsep{\fill}\hspace{10pt}}rrrrrr@{\extracolsep{0pt}}l@{\extracolsep{\fill}}}
\toprule
 & \multicolumn{7}{c}{GPT-5.6} & \multicolumn{7}{c}{Claude Opus 5} \\
\cmidrule(lr){2-8}\cmidrule(lr){9-15}
Context & Tok. & $\varepsilon$ & $\rho$ & App. & Res.$\uparrow$ & \multicolumn{2}{c}{$\Delta$} & Tok. & $\varepsilon$ & $\rho$ & App. & Res.$\uparrow$ & \multicolumn{2}{c}{$\Delta$} \\
\midrule
\textbf{Ours (CQ2-CARE)} & 13.8 & 100.0 & 79.0 & 98.0 & \textbf{43.7} &  &  & 27.1 & 100.0 & 84.0 & 83.3 & \textbf{46.7} &  &  \\
\midrule
\multicolumn{15}{@{}l}{\emph{Compressors}} \\
\quad LLMLingua-2 & 13.1 & 17.5 & 78.3 & 70.7 & 30.0 & {\boldmath\textbf{$-$13.7}} & $^{***}$ & 26.5 & 8.2 & 84.3 & 80.7 & 40.3 & {\boldmath\textbf{$-$6.3}} & $^{*}$ \\
\quad LongLLMLingua & 13.2 & 67.1 & 63.7 & 84.0 & 35.0 & {\boldmath\textbf{$-$8.7}} & $^{***}$ & 26.7 & 69.9 & 73.1 & 63.0 & 29.0 & {\boldmath\textbf{$-$17.7}} & $^{***}$ \\
\quad LongCodeZip & 13.1 & 96.8 & 72.2 & 89.7 & 37.3 & {\boldmath\textbf{$-$6.3}} & $^{**}$ & 26.2 & 96.9 & 82.9 & 73.3 & 36.0 & {\boldmath\textbf{$-$10.7}} & $^{**}$ \\
\quad SWE-Pruner$^\dagger$ & 13.4 & 91.7 & 77.4 & 93.0 & 41.7 & {\boldmath\textbf{$-$2.0}} &  & 27.0 & 92.1 & 83.9 & 73.7 & 39.3 & {\boldmath\textbf{$-$7.3}} & $^{*}$ \\
\multicolumn{15}{@{}l}{\emph{Retrievers}} \\
\quad BM25 (chunks) & 13.6 & 100.0 & 58.3 & 86.7 & 29.0 & {\boldmath\textbf{$-$14.7}} & $^{***}$ & 26.9 & 100.0 & 71.9 & 82.3 & 34.0 & {\boldmath\textbf{$-$12.7}} & $^{***}$ \\
\quad Dense (chunks) & 13.7 & 100.0 & 57.9 & 78.3 & 22.7 & {\boldmath\textbf{$-$21.0}} & $^{***}$ & 27.0 & 100.0 & 68.4 & 86.7 & 33.7 & {\boldmath\textbf{$-$13.0}} & $^{***}$ \\
\quad BM25 (files) & 12.9 & 100.0 & 49.8 & 95.7 & 34.3 & {\boldmath\textbf{$-$9.3}} & $^{***}$ & 26.3 & 100.0 & 58.2 & 80.7 & 34.0 & {\boldmath\textbf{$-$12.7}} & $^{***}$ \\
\quad SweRank$^\dagger$ & 13.3 & 100.0 & 71.7 & 97.3 & 43.0 & {\boldmath\textbf{$-$0.7}} &  & 26.5 & 100.0 & 75.0 & 89.3 & 45.0 & {\boldmath\textbf{$-$1.7}} &  \\
\midrule
{\color{gray}\textit{Gold files (upper)}} & {\color{gray}\textit{12.2}} & {\color{gray}\textit{100.0}} & {\color{gray}\textit{97.0}} & {\color{gray}\textit{97.7}} & {\color{gray}\textit{49.3}} &  &  & {\color{gray}\textit{12.2}} & {\color{gray}\textit{100.0}} & {\color{gray}\textit{97.0}} & {\color{gray}\textit{83.0}} & {\color{gray}\textit{46.3}} &  &  \\
{\color{gray}\textit{No source (lower)}} & {\color{gray}\textit{0}} & {\color{gray}\textit{n/a}} & {\color{gray}\textit{0.0}} & {\color{gray}\textit{42.0}} & {\color{gray}\textit{8.7}} &  &  & {\color{gray}\textit{0}} & {\color{gray}\textit{n/a}} & {\color{gray}\textit{0.0}} & {\color{gray}\textit{81.0}} & {\color{gray}\textit{13.0}} &  &  \\
\bottomrule
\end{tabular*}
\caption{\textbf{At the same per-issue token budget, exact whole files beat token compressors and retrieval, and lead a trained retriever of whole functions by less than two points.} Every context is fitted to the realized size $B_i$ of ours for that model on each issue (\S\ref{sec:setup}). File-level BM25 adds whole files while they fit, so it lands under $B_i$. Tok.: mean context tokens (thousands). $\varepsilon$: verbatim lines. $\rho$: gold-file recall. App.: patches that apply. Res.: resolved. All in \% except Tok. $\Delta$: the row's resolution minus ours (points), with $^{***}p<0.001$, $^{**}p<0.01$, $^{*}p<0.05$ in one-sided exact McNemar tests, Holm-adjusted within the compressors with SWE-Pruner and within the planned retrievers with SweRank. Bold: best resolution per model, and every $\Delta$. $^\dagger$: added after the main results, as was file-level BM25, whose tests are unadjusted.}
\label{tab:main}
\end{table*}
\begin{algorithm}[t]
\caption{\ours{}: exact-file selection}
\label{alg:ours}
\linespread{1.2}\selectfont
\begin{algorithmic}[1]
\Require issue $q$, repository $R$, model $\pi$
\Ensure context $m$; budget $B_i = |m|$
\State $P \gets \mathrm{paths}(R)$ \Comment{inventory}
\State $I \gets \mathrm{sort}_{\mathrm{alpha}}\big(\mathrm{Top}_{\tau}(P)\big)$
\State $L_0, L_1, L_2 \sim \pi_{\mathrm{loc}}(q, I)$ \Comment{localize}
\State $F \gets (L_1 \cap L_2 \cap P) \setminus P_{\mathrm{test}}$ \Comment{select}
\State \textbf{if} $F = \emptyset$ \textbf{then} $F \gets L_0 \cap P$
\State sort $F$ by $\sum_{j=1}^{2} 1/\mathrm{rank}_{L_j}(f)$, descending
\State $Q \gets \pi_{\mathrm{plan}}(q, F)$, \ $|Q| \le 4$; \ $E \gets (\,)$ \Comment{evidence}
\State \textbf{for} $s \in Q$ \textbf{do}
\State \hspace{\algorithmicindent}$G \gets \mathrm{grep}(R, s) \setminus \big(F \cup \mathrm{files}(E)\big)$
\State \hspace{\algorithmicindent}\textbf{for} $g \in \mathrm{Top}_2(G)$ \textbf{while} $|\mathrm{files}(E)| < 8$ \textbf{do}
\State \hspace{\algorithmicindent}\hspace{\algorithmicindent}$E \gets E \oplus \mathrm{span}(g, s)$
\State $m \gets (\,)$ \Comment{assemble}
\State \textbf{for} $b \in \mathrm{cut}_{60\mathrm{K}}(F_{\mathrm{code}}) \oplus \mathrm{cut}_{12\mathrm{K}}(E)$ \textbf{do}
\State \hspace{\algorithmicindent}\textbf{if} $|m \oplus b| \le 80\mathrm{K}$ \textbf{then} $m \gets m \oplus b$
\State \Return $m$
\end{algorithmic}
\end{algorithm}
\emph{Exact-file selection} predicts the files $F$ the fix must touch and shows them whole and verbatim (up to size caps, Algorithm~\ref{alg:ours}), followed by short evidence spans $E$ from other files ($\oplus$ concatenates):
\begin{equation}
  m_{\mathrm{sel}} = \big(f[1{:}|f|]\big)_{f \in F} \oplus E.
  \label{eq:sel}
\end{equation}

What each context keeps is described by two quantities, both measured offline:
\begin{align}
  \rho(m) &= \frac{|F^\star \cap \{f_1, \dots, f_K\}|}{|F^\star|}, \label{eq:rho}\\
  \varepsilon(m) &= \frac{|\{\ell \in m : \ell \in f(\ell)\}|}{|\{\ell \in m\}|}. \label{eq:eps}
\end{align}
The \emph{gold-file recall} $\rho(m)$ is the share of $F^\star$, the pre-existing files modified by the gold patch, that appear in at least one span of $m$; it uses the gold patch, so we compute it after the fact and never show it to a model.
The share of \emph{verbatim lines} $\varepsilon(m)$ runs over the nonblank context lines $\ell$ and counts those that equal, up to trailing whitespace, some line of $f(\ell)$, the file their header names.
Retrieval \eqref{eq:ret} keeps $\varepsilon = 1$ and can lose $\rho$; token pruning \eqref{eq:prune} of the right files keeps $\rho$ and lowers $\varepsilon$; exact-file selection \eqref{eq:sel} keeps $\varepsilon = 1$, except where a size cap cuts a line ($\varepsilon \ge 0.996$), and spends the budget on $\rho$.

Our selector, \ours{}, instantiates \eqref{eq:sel} with the model under test, $\pi$, making every call (Algorithm~\ref{alg:ours}).
The model never sees the whole repository, only a path inventory $I$: the paths $P$ of $R$ that share the most words with $q$, up to $\tau = \tfrac{1}{2}\min(|P|, 32\mathrm{K})$ tokens, listed alphabetically ($\mathrm{Top}_k$ keeps the $k$ best-ranked items, and $|\cdot|$ is a size in tokens; lines~1--2).
Given $q$ and $I$, the model returns independent ranked answers $L_0, L_1, L_2 \sim \pi_{\mathrm{loc}}(q, I)$ to ``which files must change?'' (line~3).
The selected files $F$ are those that both $L_1$ and $L_2$ name, outside the paths $P_{\mathrm{test}}$ under test and documentation directories, taking agreement between samples as a sign of reliability \citep{wang2023selfconsistency}; if none remain (a malformed answer counts as empty), $F$ falls back on $L_0$ (lines~4--6).
The model then reads $q$ and $F$ and requests at most four typed searches $Q = \pi_{\mathrm{plan}}(q, F)$, such as a definition or its callers; for each search $s$, $\mathrm{grep}(R, s)$ finds the files that contain a fixed string derived from $s$, and for at most two of them not yet shown, $g$, the evidence $E$ gains $\mathrm{span}(g, s)$, the enclosing definition of the match or a window around it, until $E$ covers eight files ($|\mathrm{files}(E)| = 8$; lines~7--11).
The context concatenates the source-code files $F_{\mathrm{code}}$ in full and the spans of $E$, where $\mathrm{cut}_c$ keeps the first $c$ tokens of each block and a block that would take $m$ past 80K tokens is skipped (lines~12--14).

The budget is set per issue by our exact-file selector, which itself takes no budget: we build its context for issue $i$ first, and its realized token count $B_i = |m_{\mathrm{sel}}|$ (o200k tokenizer, headers included) is the budget for every compressor and retriever on that issue.
Retrievers pack up to $B_i$; each compressor is fitted to $B_i$ within $\pm$2\% by searching only its own rate setting $r$, and when no setting lands in the window, the closest result not above $1.02\,B_i$ is used (in one LongCodeZip context none was, and the smallest, $1.97\,B_i$, is kept).

\section{Experiments}
\label{sec:exp}

\subsection{Data}
\label{sec:issues}

We use the SWE-rebench leaderboard \citep{badertdinov2025swerebench}, which collects recent issues automatically, at a pinned revision with 860 Python tasks from 413 repositories; on the older SWE-bench Verified, models can often name the files to edit from the issue text alone, which its authors attribute in part to possible memorization \citep{liang2026sweillusion}.
A salted hash drew 340 candidate issues, one per repository, before any model call.
Each candidate passed an eligibility check in our harness: the gold patch resolves it and an empty patch does not (328 of 340 passed).
The first 300 eligible issues form the population, created between January 2025 and May 2026; a hash-chosen subset of 100 is used for the rate sweep and the edit-format control.

\subsection{Setup and Baselines}
\label{sec:contexts}

All methods share one pipeline and differ only in the context step (Appendix~\ref{app:impl}).
\ours{} (\S\ref{sec:setup}) draws five ranked answers per issue, of which a salted hash picks $L_1$, $L_2$ and $L_0$; with GPT-5.6, its context holds 1.7 selected files on average (median 1) and spans from 6.0 more files (variants in \S\ref{sec:selector}).

\begin{figure*}[t]
\centering
\figsafe{width=\textwidth}{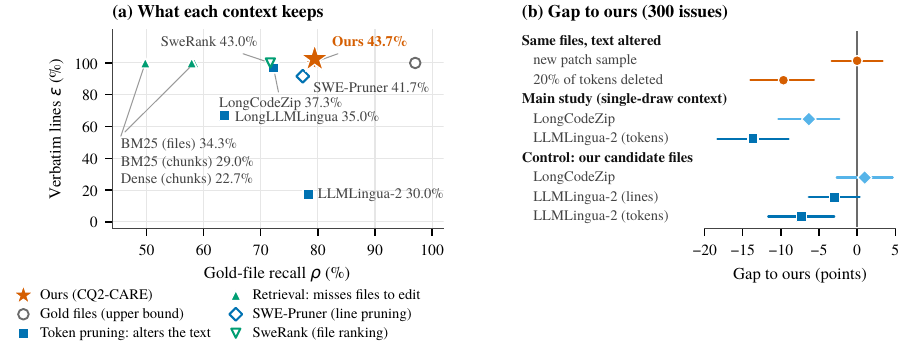}
\caption{\textbf{Why compressed contexts fail.} (a) Gold-file recall $\rho$ and verbatim lines $\varepsilon$ of every context, labelled with its resolution (GPT-5.6, 300 issues, budgets matched per issue; the gold files resolve 49.3\%)\rev{. Hollow markers: the two trained methods, SWE-Pruner and SweRank}. (b) Gap to ours, the paired difference in resolution (row minus ours, points), on all 300 issues: a second patch sample; our contexts with 20\% of every line's tokens deleted; the main-study compressors, which cut the larger single-draw context; and, as a control rather than a baseline, the same compressors cutting our own candidate files to $B_i$. Bars: 95\% issue-bootstrap intervals.}
\label{fig:overview}
\end{figure*}

The baselines are context steps an agent builder can drop in without training.
Each compressor cuts $\tilde m$, the uncompressed context that the same pipeline builds from $L_0$ alone, to $B_i$ and keeps span headers verbatim ($\tilde m$ already fits in 52 issues and is used as is).
We run the official implementations with their documented settings: LLMLingua-2 \citep{pan2024llmlingua2}, LongLLMLingua \citep{jiang2024longllmlingua} with Qwen2.5-7B-Instruct \citep{qwen2025qwen25} as the scorer, and LongCodeZip \citep{shi2025longcodezip} with Qwen2.5-Coder-7B-Instruct \citep{hui2024qwen25coder}.

Chunk BM25 \citep{robertson2009bm25} and dense retrieval with CodeRankEmbed \citep{suresh2025cornstack} rank top-level Python definitions or 60-line windows against the issue and pack them in rank order up to $B_i$.
File-level BM25, SWE-bench's own retrieval baseline \citep{jimenez2024swebench}, adds whole non-test Python files in rank order while they fit, so it lands under $B_i$.
\rev{Two methods trained for issue resolution were added after the main results, with the same prompt and grader and without tests, gold patches or fault locations (Algorithms~\ref{alg:swepruner} and~\ref{alg:swerank}): SWE-Pruner \citep{wang2026swepruner} deletes lines from the single-draw context that the compressors cut at a threshold that fits $B_i$, and SweRank \citep{reddy2026swerank} ranks every non-test Python function against the issue; we show whole functions in that order up to $B_i$.}
For scale, the gold-file oracle (upper bound) shows the source files the gold patch modifies, and the no-source context shows the issue alone.

The main table reports one model per family, each used for every call: \texttt{gpt-5.6-sol} (medium reasoning effort) and Claude Opus 5, both on all 300 issues.
The model returns a unified diff from a prompt that differs across contexts only in the localized paths and the context, and we grade it with the SWE-rebench fork of the official harness \citep{jimenez2024swebench,badertdinov2025swerebench}; patches that fail to parse or apply count as unresolved (exit codes alone are not enough; Appendix~\ref{app:tacit}).
Comparisons are paired by issue: one-sided exact McNemar tests \citep{mcnemar1947,dror2018hitchhiker}, \rev{Holm-corrected \citep{holm1979} within the compressors, with SWE-Pruner, and within the planned retrievers, with SweRank (Appendix Table~\ref{tab:hyp}),} with 95\% issue-bootstrap intervals \citep{efron1994bootstrap}.
The main tests follow a written plan, which we did not time-stamp; file-level BM25 and the analyses of \S\ref{sec:analysis} are post hoc (Appendix~\ref{app:posthoc}).

\subsection{Main Results}
\label{sec:main}

Exact files beat every drop-in baseline with both models (Table~\ref{tab:main}).
\ours{} resolves 43.7\% of the issues with GPT-5.6 and 46.7\% with Claude Opus 5, 6.3 points more than the best compressor with either model and 9.3 and 12.7 points more than the best retriever.
All these leads are significant, the narrowest (LLMLingua-2 with Claude, $p = 0.027$) only weakly.
Whole files alone are not enough: SWE-bench's file-level BM25 shows whole files chosen by similarity to the issue and trails by 9.3 and 12.7 points.
\rev{\ours{} even resolves more issues than the two trained methods: SWE-Pruner trails with both models, significantly only with Claude, and SweRank is within two points (\S\ref{sec:lose}).}

Exact files chosen by the model come close to the true ones: the gold files resolve 5.7 points more than \ours{} with GPT-5.6 and no more with Claude.
At the budget a compressor would use, an agent that chooses a few whole files recovers most of what knowing the files to edit would give it.

\section{Analysis}
\label{sec:analysis}

\subsection{Retrievers Miss Files, Compressors Alter Text}
\label{sec:lose}

\begin{table}[t]
\centering
\footnotesize
\setlength{\tabcolsep}{3pt}
\begin{tabular}{lrrrrr}
\toprule
 & \multicolumn{4}{c}{Lines of the context (\%)} & \\
\cmidrule(lr){2-5}
Context & To edit & Other & Tests & Non-Py. & Miss \\
\midrule
\textbf{Ours} & 51.7 & 31.4 & 8.8 & 8.1 & 3.4 \\
BM25 (chunks) & 20.9 & 34.0 & 17.2 & 28.0 & 26.2 \\
Dense (chunks) & 16.5 & 25.6 & 31.6 & 26.4 & 27.5 \\
BM25 (files) & 41.8 & 57.6 & 0.0 & 0.0 & 29.2 \\
\bottomrule
\end{tabular}
\caption{\textbf{Retrievers often miss the files to edit.} Share of each context's lines by the file they come from, averaged over issues (GPT-5.6; the 298 issues whose gold patch modifies an existing file): to edit: files the gold patch modifies; other: other Python source; non-Py.: non-Python files such as documentation and templates. Miss: issues (\%) whose context shows no file to edit. File-level BM25 ranks only non-test Python files.}
\label{tab:where}
\end{table}

\begin{table}[t]
\centering
\footnotesize
\setlength{\tabcolsep}{3pt}
\begin{tabular}{@{}l>{\raggedright\arraybackslash}p{4.4cm}r@{}}
\toprule
Context & What fills the budget & Res. \\
\midrule
BM25 (chunks) & 7 test files, e.g.\ \texttt{test\_ignore.py} & no \\
Dense (chunks) & 7 test files, e.g.\ \texttt{test\_diff.py} & no \\
BM25 (files) & \texttt{cli/parser.py}, \texttt{commands/diff.py} & no \\
\textbf{Ours} & \texttt{repo/diff.py} whole, callers, 2 tests & yes \\
\bottomrule
\end{tabular}
\caption{\textbf{Retrievers rank what echoes the issue} (\texttt{iterative\_\_dvc-10844}, GPT-5.6). The issue is a shell script that runs \texttt{dvc add} and \texttt{dvc diff}; the fix edits \texttt{dvc/repo/diff.py}, which no retriever shows. Chosen for illustration from the 12 issues where ours resolves and both chunk retrievers fail without showing any file to edit.}
\label{tab:case}
\end{table}

Retrievers lose the files to edit (Figure~\ref{fig:overview}a).
Ranking code by its similarity to the issue fills the budget with other code, tests and documentation (Table~\ref{tab:where}): chunk BM25 and dense retrieval put a fifth or less of their lines in the files to edit, against half for \ours{}, and show none of them on over a quarter of the issues.
On an issue that reports a shell script (Table~\ref{tab:case}), chunk BM25 fills the budget with tests that share its commands and dense retrieval with tests of the \texttt{diff} command, while the file to edit appears in neither.
File-level BM25 shows whole files, so its patches apply more often and it resolves more than chunk BM25, but most of its lines still come from files the fix does not touch.
\rev{SweRank, trained to rank functions for an issue, comes closest, 0.7 and 1.7 points behind, by showing whole functions verbatim, so its patches apply (97.3\% and 89.3\%), but it shows every line the gold patch changes on only 39\% of the GPT-5.6 issues, against 70\% for \ours{} (Appendix Table~\ref{tab:cmp}).}

Token compressors lose the verbatim text, and their patches stop applying.
The fewer lines a compressor keeps verbatim, the fewer patches apply: from LongCodeZip (96.8\% of lines verbatim, 89.7\% of patches apply) to LLMLingua-2 (17.5\% and 70.7\%), against 98.0\% for \ours{}.
Patches that do not apply account for 28 of the 48 issues that \ours{} resolves and LLMLingua-2 does not; where both patches apply, a smaller gap remains, consistent with altered text also misleading the edit itself.
\rev{SWE-Pruner, trained to prune lines for coding agents, keeps 91.7\% of its lines verbatim and still trails by 2.0 and 7.3 points, because the gaps it leaves inside files cost patches: 93.0\% apply with GPT-5.6 and 73.7\% with Claude, against 98.0\% and 83.3\% for \ours{}.}

\subsection{Controlled Tests}
\label{sec:controlled}


The comparison above is observational: the compressors differ in which files they keep as well as in how they alter them.
Two post-hoc tests on all 300 issues change one factor at a time \rev{(Figure~\ref{fig:overview}b; Appendix Table~\ref{tab:controlled}).}

Altering the text alone, with the files fixed, costs about ten points.
Deleting 20\% of the tokens of every line of our own contexts lowers patch applicability from 98.0\% to 78.7\% and resolution by 9.7 points ($p < 0.001$), while a second patch sample of the intact contexts resolves exactly as many issues.

Given the same files, a compressor that cuts whole lines matches \ours{}.
This test is a control, not a competing baseline: the compressors cut our own candidate files, built by our localization and evidence step, so they inherit our file choice and only the cut differs.
When they cut these files to $B_i$ instead of the larger single-draw context, LongCodeZip resolves 44.7\%, against 43.7\% for \ours{}; LLMLingua-2 comes within 3.0 points when its scores choose whole lines, but trails by 7.3 when they choose tokens.
\ours{}'s lead over LongCodeZip in the main comparison therefore comes from which files it keeps: cutting the single-draw context lost files to edit that cutting our candidates kept.
A within-issue regression over all (issue, context) pairs agrees: verbatim lines and gold-file recall both raise resolution, and the effect of verbatim lines runs mostly through whether the patch applies (Appendix Table~\ref{tab:w2}b).
\rev{An estimator built on these quantities predicts the resolution of each main context left out of its fit (Appendix~\ref{app:newbaselines}).}

\begin{table}[t]
\centering
\footnotesize
\setlength{\tabcolsep}{4pt}
\begin{tabular}{lr@{}lr@{}lr@{}l}
\toprule
 & \multicolumn{6}{c}{Gap to ours (points)} \\ \cmidrule(l){2-7}
 & \multicolumn{4}{c}{In tolerance} & \multicolumn{2}{c}{Search/replace} \\ \cmidrule(lr){2-5}\cmidrule(l){6-7}
Baseline & \multicolumn{2}{c}{GPT-5.6} & \multicolumn{2}{c}{Claude} & \multicolumn{2}{c}{GPT-5.6} \\
\midrule
LLMLingua-2 & $-$16.7 & $^{***}$ & $-$6.6 & $^{*}$ & $-$15.0 & $^{***}$ \\
LongLLMLingua & $-$11.6 & $^{***}$ & $-$19.8 & $^{***}$ & $-$13.0 & $^{**}$ \\
LongCodeZip & $-$6.6 & $^{*}$ & $-$9.6 & $^{*}$ & $-$9.0 & $^{*}$ \\
SWE-Pruner$^\dagger$ & $-$1.6 &  & $-$7.0 & $^{*}$ & \multicolumn{2}{c}{--} \\
\midrule
BM25 (chunks) & $-$14.6 & $^{***}$ & $-$13.2 & $^{***}$ & $-$19.0 & $^{***}$ \\
Dense (chunks) & $-$20.9 & $^{***}$ & $-$12.8 & $^{***}$ & $-$17.0 & $^{***}$ \\
SweRank$^\dagger$ & $-$1.0 &  & $-$1.1 &  & \multicolumn{2}{c}{--} \\
\bottomrule
\end{tabular}
\caption{\textbf{The drop-in baselines still trail ours in two further checks.} Gap to ours: the baseline's resolution minus ours. In tolerance: only the issues where the baseline's realized tokens are within $\pm$2\% of $B_i$ (197--297 of 300 per baseline and model). Search/replace: both contexts with exact-match search/replace edits instead of diffs (GPT-5.6, 100 issues, not run for the two trained methods). Stars as in Table~\ref{tab:main}. $^\dagger$: added after the main results.}
\label{tab:robust}
\end{table}

\begin{figure}[t]
\centering
\figsafe{width=\columnwidth}{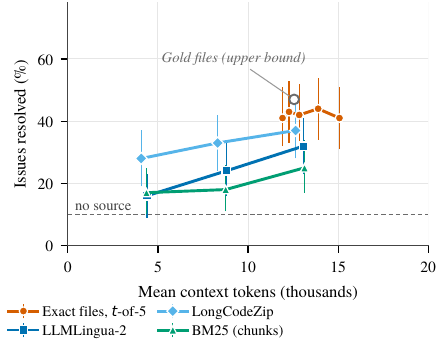}
\caption{\textbf{The ordering holds across budgets.} Resolution vs.\ context size on 100 issues (GPT-5.6): compressors and chunk BM25 at 25/50/75\% of the uncompressed context, and exact-file contexts from $t$-of-5 localization agreement. Bars: 95\% issue-bootstrap intervals.}
\label{fig:budgets}
\end{figure}

\subsection{Robustness}
\label{sec:robust}

The baselines still trail ours in four further checks with GPT-5.6 \rev{(Table~\ref{tab:robust}, Figure~\ref{fig:budgets}; details in Appendix~\ref{app:analysis})}.
\emph{Budget tolerance:} the compressors miss the $\pm$2\% window on a quarter to a third of the issues, almost always by landing under budget; on the issues within it, every gap is unchanged or larger.
\emph{Edit format:} with search/replace edits instead of diffs, every baseline still trails ours, and quotes taken from compressed contexts often match no unique location in the real file (on 45 issues for LLMLingua-2, against 4 for \ours{}).
\emph{Compression rate:} at 25\%, 50\% and 75\% of the uncompressed context, \rev{no compressor or retriever reaches any exact-file context (Figure~\ref{fig:budgets}).}
\emph{Patch sampling:} a second patch sample moves the rates of \ours{}, LongCodeZip and LLMLingua-2 by at most 2.0 points, and averaged over both samples \ours{} still leads both ($p \le 0.002$).

\subsection{Ablations of \ours{}}
\label{sec:selector}

What makes exact-file selection work is predicting few files, not agreement between samples (Appendix Table~\ref{tab:variants}).
Agreement raises localization F1 well above a single draw (0.613 vs.\ 0.430), but only because it names half as many files: a single draw cut to the same number of files reaches 0.600 and resolves about as many issues with both models (Claude: Appendix Table~\ref{tab:claude}).
More exact files help when the budget allows: the files that either draw names resolve 4.0 points more than \ours{} at 2.3K more tokens, and the agreement thresholds $t$-of-5 \rev{trade files for tokens smoothly (Figure~\ref{fig:budgets}).}
The evidence step adds 3.0 points, short of significance ($p = 0.068$).

Exact-file selection is also the cheaper way to fill the budget.
Against the uncompressed single draw, \ours{} uses 0.68$\times$ the tokens with GPT-5.6 and passes the planned non-inferiority test; with Claude, it uses 0.73$\times$ the tokens and resolves more, but the token ratio's upper bound narrowly misses its margin (Appendix~\ref{app:posthoc}).
Its whole pipeline costs \$0.151 per issue with GPT-5.6, against \$0.199 for the single draw and \$0.168--0.177 for the compressor contexts in API calls alone, because those first build the uncompressed context.

\section{Discussion}
\label{sec:discussion}

\paragraph{Choose files or compress tokens?}
An agent builder who must cut context can retrieve chunks, prune a larger context, or choose files and show them whole.
Retrieval keeps the text but often loses the files to edit, and token pruning keeps the files but alters the text that an edit must quote; each costs on its own (\S\ref{sec:controlled}).
Edits stay locatable only when the text is verbatim (\S\ref{sec:robust}), and editing tools that replace a quoted span \citep{anthropic2024swebench,aider-repo} or a numbered line range \citep{yang2024sweagent,wang2025openhands} need exactly this.
Choosing files also needs no compression model: LongLLMLingua and LongCodeZip spent a median 34--37\,s of GPU time per issue (84--97\,s on Claude's larger contexts, where LongCodeZip ran out of memory on 4 issues at 48\,GB), and \ours{} adds one localization call.
What the results do not show is that a selector beats a code-aware compressor that starts from the same files: given our candidates, LongCodeZip matched \ours{}, although it still spent 6.8 GPU-hours on 300 issues.
We read this as a rule rather than a ranking: choose the candidate files well, then cut whole lines if anything must be cut.

\paragraph{Re-read on demand or compress once?}
Agents that can re-open files may recover some of the losses we measure, because a compressed view tells them which file to read in full.
Each re-read, however, costs a turn and the tokens that compression was meant to save, and a model may not notice that the text it sees is altered: in Figure~\ref{fig:main}, it rewrote a line that does not exist in the file.
Measuring this trade-off needs a full agent loop, which we leave to future work.

\paragraph{Localization scores or end-to-end tests?}
Intermediate metrics can mislead.
Agreement between two localization samples raises file-level F1 far above a single sample, yet a single sample cut to the same number of files resolves about as many issues (\S\ref{sec:selector}).
Exit-code grading, which we used before this study, credited an empty patch on 4 of 6 issues (Appendix~\ref{app:tacit}).
Context methods for coding agents should therefore be compared end to end, at equal budgets, with the benchmark's tests.

\section{Related Work}
\label{sec:related}

\paragraph{Prompt compression.}
Hard prompt compression removes tokens or sentences that a smaller model judges uninformative \citep{li2023selectivecontext,jiang2023llmlingua,jiang2024longllmlingua,pan2024llmlingua2}; RECOMP compresses retrieved documents extractively or abstractively \citep{xu2024recomp}, and \citet{li2025promptcompressionsurvey} survey the area.
On long-context question answering and summarization, \citet{jha2024characterizing} find that extractive, reranker-based selection of chunks often outperforms token pruning and abstractive summarization.
We find the same ordering for issue resolution and identify the cost specific to it: edits must quote the text that pruning alters.

\paragraph{Agent memory.}
Language-agent frameworks separate a working memory, from which each prompt is built, from long-term memories retrieved into it \citep{sumers2024coala}, and MemGPT manages what moves between a limited context and external storage \citep{packer2023memgpt}; for a coding agent, the repository is the long-term memory, and the question is which part of it enters the prompt.
For code, LongCodeZip keeps whole functions and blocks chosen by perplexity, since its authors note that token-level compression ``corrupts code syntax and structure'' \citep{shi2025longcodezip}, and perplexity-based pruning removes Python indentation in repository-level completion \citep{feng2026repocompression}.
SWEzze trains a compressor on execution-derived minimal contexts, but compares methods only around fault locations already found by Agentless \citep{jia2026swezze}; SWE-Pruner prunes the lines an agent reads and attributes, as a hypothesis, the losses of LLMLingua-2 and embedding retrieval in a 50-issue comparison to disrupted syntax and missed details \citep{wang2026swepruner}. \rev{SweRank learns to retrieve and rerank whole functions from issue--patch pairs \citep{reddy2026swerank}.}
We instead fix the budget per issue, let each method decide which files enter the context, and measure both failure modes with controlled tests.

\section{Conclusion}
\label{sec:conclusion}

We compared ways to fill a coding agent's context budget at equal per-issue budgets on 300 test-graded issues with two model families.
\rev{Showing a few whole files verbatim resolved more issues than pruning tokens or retrieving chunks, and about as many as a trained retriever of whole functions, because} retrieval often misses the files to edit and token pruning alters the text that patches must quote, and controlled tests showed that each loss costs on its own.
For coding agents, the context budget is best spent choosing which code to show, kept verbatim and in whole lines.

\section*{Limitations}

\begin{itemize}[leftmargin=*,itemsep=1pt,topsep=2pt]
\item \textbf{Setting.} We study one patch call per context on Python issues from one benchmark, not a multi-turn agent that could read more files or run tests.
\item \textbf{Comparisons.} $B_i$ is our context's realized size, so we compare at our method's operating point (the 100-issue rate sweep covers others). File-level BM25, the controlled tests, SWE-Pruner (in its single-turn mode) and SweRank were added after the main results; SweRank may have trained on other pull requests from our repositories, and its small gap to \ours{} does not show equal performance.
\item \textbf{Estimator.} It uses the gold patch and serves only as a diagnostic; its coverage counted inside functions was chosen after seeing SweRank's result and needs a test on an unseen method.
\item \textbf{Precision and contamination.} Most cells have one patch sample, and a second one flips 8.7--18.9\% of outcomes, so small gaps such as Claude's to LLMLingua-2 ($p = 0.027$) are fragile. The issues may overlap the models' training data, but the gaps do not shrink detectably on newer issues (Appendix Table~\ref{tab:w2}a); some may come from repositories seen while developing \ours{}.
\end{itemize}

\section*{Ethics Statement}
This work evaluates context-compression methods on public open-source repositories and issues. It involves no human subjects or personal data. Generated patches were executed only inside isolated sandboxes. We will release code, prompts and per-issue outcomes.

\section*{Acknowledgments}
We thank the INVITE Institute at the University of Illinois Urbana-Champaign and the ACCESS program for the computing resources that supported this work.
This material is based upon work supported by the National Science Foundation and the Institute of Education Sciences under Grant \#2229612. Any opinions, findings, and conclusions or recommendations expressed in this material are those of the author(s) and do not necessarily reflect the views of National Science Foundation or the U.S. Department of Education.
This work used the Delta system at the National Center for Supercomputing Applications [award OAC 2005572] through allocations CIS251094 and CIS261897 from the Advanced Cyberinfrastructure Coordination Ecosystem: Services \& Support (ACCESS) program \citep{boerner2023access}, which is supported by National Science Foundation grants \#2138259, \#2138286, \#2138307, \#2137603, and \#2138296.


\appendix

\section{A Random Example}
\label{app:example}

Figure~\ref{fig:main} shows an issue chosen for illustration.
To show what each compressor does to code without choosing a flattering case, we also drew one issue by a salted hash (the first issue in hash order whose \ours{} context and all three compressor contexts share a non-test Python file) and print the first lines of that file's span in each context.
The issue is \texttt{scikit-hep\_\_cabinetry-506}.
LongLLMLingua and LongCodeZip keep this region verbatim. LLMLingua-2 drops keywords and punctuation (the \texttt{import} before \texttt{pathlib}, \texttt{import} and the commas in the \texttt{typing} line, \texttt{cabinetry import} before \texttt{route}) and changes leading whitespace, so most of these lines no longer match the file exactly.

\noindent File: \path{src/cabinetry/templates/collector.py}

\noindent\textbf{CQ2-CARE (ours)}, first 9 lines of the span
\begin{Verbatim}[fontsize=\scriptsize,frame=single]
"""Collects required template histograms provided by user."""

import logging
import pathlib
from typing import Any, cast, Dict, Literal, Optional

from cabinetry import histo
from cabinetry import route
from cabinetry.templates import utils
\end{Verbatim}

\noindent\textbf{LLMLingua-2}, first 9 lines of the span
\begin{Verbatim}[fontsize=\scriptsize,frame=single]
"Collects required template histograms provided by user."

 import logging
 pathlib
 from typing Any, cast Dict Literal Optional

 from cabinetry import histo
 from route
 from cabinetry.templates import utils
\end{Verbatim}

\noindent\textbf{LongLLMLingua}, first 9 lines of the span
\begin{Verbatim}[fontsize=\scriptsize,frame=single]
"""Collects required template histograms provided by user."""

import logging
import pathlib
from typing import Any, cast, Dict, Literal, Optional

from cabinetry import histo
from cabinetry import route
from cabinetry.templates import utils
\end{Verbatim}

\noindent\textbf{LongCodeZip}, first 9 lines of the span
\begin{Verbatim}[fontsize=\scriptsize,frame=single]
"""Collects required template histograms provided by user."""

import logging
import pathlib
from typing import Any, cast, Dict, Literal, Optional

from cabinetry import histo
from cabinetry import route
from cabinetry.templates import utils
\end{Verbatim}

\section{Implementation Details and Compute}
\label{app:impl}

This appendix gives the details of \ours{} (Algorithm~\ref{alg:ours}) and of the baselines. \rev{Algorithms~\ref{alg:fit}--\ref{alg:bounds} give every other context step in the same notation, with a link to each method's paper and code, including the two trained methods (Algorithms~\ref{alg:swepruner} and~\ref{alg:swerank}).}

\paragraph{Localization and evidence.}
The path inventory ranks paths by $h$, the number of distinct words they share with the issue text and the fixed question the model is asked, with ties broken by $h/n$, the share of the path's $n$ words that match; words are lowercased identifiers and their parts split at punctuation, underscores and case changes, so the example path \texttt{path/to/file.py} in that question makes \texttt{py} a shared word for every Python file. The ranking only decides which paths fit, counting bytes/3 as tokens; the model sees them in alphabetical order, and the list omits files under \texttt{node\_modules}, \texttt{vendor}, \texttt{dist}, \texttt{build}, \texttt{target} and \texttt{.next} directories. A salted hash assigns the roles of the five answers drawn per issue: two construction samples $L_1, L_2$, one comparator $L_0$, and two that serve only the agreement-threshold sweep. \ours{} keeps the files both construction samples name if they exist in the repository, even if they were cut from the inventory, drops paths under \texttt{test(s)}/\texttt{doc(s)} directories and orders the rest by summed reciprocal rank; it falls back on $L_0$ when a sample is malformed or nothing remains (1.7\% of issues with GPT-5.6, 0.3\% with Claude). Each evidence search (e.g., a definition, callers, a literal string, related tests) becomes up to eight fixed strings for \texttt{git grep} at the base commit and adds a span from each of at most two new files, eight in total: the enclosing Python definition of the match or a window of lines around it. A selected file over 60K tokens is cut to its first 60K, selected non-code files are not shown (13 files), evidence spans are capped at 12K tokens each, and any block that would take the context past 80K tokens is skipped; with GPT-5.6, 7 of 300 contexts reach a cap.

\paragraph{Models and calls.}
Every call uses the same model per run: \texttt{gpt-5.6-sol} through the Azure OpenAI Responses API, or Claude Opus 5 through Microsoft Foundry, with medium reasoning effort and the providers' default sampling. Output limits are 6,144 tokens for localization, 4,096 for the planner and 32,768 for patches. Requests are byte-identical across contexts except for the context; failed transport calls are retried, and refusals or truncated answers count as unresolved. Localization returns a ranked JSON list of paths; the planner returns at most four typed search requests; the patch call returns a unified diff (or, in the edit-format control, search/replace edits, located by exact and then whitespace-tolerant matching) in a JSON field. Post-processing, identical for every context, fixes diff headers and hunk line counts and re-anchors a hunk only when its original lines match the base-commit source exactly and uniquely; it never reads tests or the gold patch.

\paragraph{Contexts.}
Token counts use the o200k tokenizer. Repository names, instance identifiers and commit hashes are masked in everything the model sees. Compressors search only their own rate setting for the closest result within $\pm$2\% of $B_i$ (at most 16 settings per issue) and keep span headers verbatim. LLMLingua-2 uses the \texttt{xlm-roberta-large} MeetingBank classifier and keeps the highest-scored words within each chunk of up to 512 tokens; LongLLMLingua takes the issue as the question, drops and reorders whole spans, then prunes tokens at a rate set per span; LongCodeZip, with the settings of its paper, keeps the top-ranked functions and chooses blocks within them by a knapsack. Chunk BM25 uses $k_1{=}1.2$, $b{=}0.75$; chunk BM25 and dense retrieval rank top-level Python definitions and the code between them, or 60-line windows for other files and for definitions over 4K tokens, and cut the first unit that does not fit to its longest whole-line prefix before stopping; CodeRankEmbed uses cosine similarity, batch size 8 and a 4,096-token limit. File-level BM25 reproduces SWE-bench's \texttt{make\_datasets} procedure (\texttt{swebench} 4.0.3): every Python file whose path does not contain the word \texttt{test}, \texttt{tests} or \texttt{testing} is one document (its path followed by its contents), BM25 uses Pyserini's default $k_1{=}0.9$, $b{=}0.4$ with our code tokenizer, and the top 20 files are added whole in rank order, skipping any file that does not fit in $B_i$. LongLLMLingua and LongCodeZip ran on RTX A6000 GPUs, LLMLingua-2 on CPU.

\paragraph{Grading and statistics.}
The SWE-rebench fork of the SWE-bench harness (\texttt{swebench} 4.0.3) runs in Apptainer sandboxes built from the benchmark's images, with a fresh overlay per context and a 30-minute limit per evaluation. Of the 340 drawn candidates, 10 gold patches failed and 2 empty patches passed the eligibility check. Bootstrap intervals use 20,000 issue resamples with seed 20260927; McNemar tests are exact; missing outcomes count as unresolved.

\paragraph{Hardware and cost.}
Grading ran on CPU cores of a shared GPU cluster; LongLLMLingua, LongCodeZip and dense retrieval ran on RTX A6000 (48\,GB) and A100 (40\,GB) GPUs, and LLMLingua-2 on CPU. For the GPT-5.6 study they used 6.5, 8.3 and 3.3 GPU-hours, and LLMLingua-2 1.3 CPU-hours; for Claude's larger contexts, LongLLMLingua and LongCodeZip used 14.2 and 11.4 GPU-hours. LongCodeZip on our candidate files used 6.8 GPU-hours. API cost at 2026 list prices was about \$830 for GPT-5.6 (including \$43 for the 100-issue controlled tests, \$179 for the 300-issue controlled tests and the second patch sample, \$23 for the search/replace runs of three more baselines and \$27 for file-level BM25) and \$1,438 for Claude Opus 5, including \$35 for its gold-file oracle on the 200 issues the first run had not covered, \$93 for file-level BM25 and about \$170 of duplicate patch calls from a pipeline error, which we graded as a second patch sample.


\begin{algorithm}[t]
\caption{\textbf{Fitting a compressor to the budget} (shared by LLMLingua-2, LongLLMLingua and LongCodeZip; SWE-Pruner searches its threshold the same way). $\theta$ is the method's own rate setting; headers count toward the budget.}
\label{alg:fit}
\small
\begin{algorithmic}[1]
\Require uncompressed context $S$ (spans with headers), issue $q$, budget $B_i$, compressor $C_\theta$, monotone grid $\Theta$ of settings
\Ensure context $m$ with $|m| \approx B_i$
\If{$|S| \le B_i$} \Return $S$ \EndIf
\State $\theta \gets$ the grid point nearest to $(B_i - |\text{headers}|)/(|S| - |\text{headers}|)$
\For{at most 16 evaluations}
  \State $m_\theta \gets$ render each span's header verbatim, followed by $C_\theta$'s output for its body; drop spans emptied by $C_\theta$
  \If{$(1-0.02)B_i \le |m_\theta| \le (1+0.02)B_i$} \Return $m_\theta$ \EndIf
  \State move $\theta$ by interpolation on $(\theta, |m_\theta|)$, with bisection as a safeguard
\EndFor
\State \Return the $m_\theta$ closest to $B_i$ that is not above $1.02\,B_i$ (the miss is recorded)
\end{algorithmic}
\end{algorithm}

\begin{algorithm}[t]
\caption{\textbf{LLMLingua-2} \citep{pan2024llmlingua2}: token classification, budget-matched with Algorithm~\ref{alg:fit}. Paper: \href{https://arxiv.org/abs/2403.12968}{arXiv:2403.12968}; code: \href{https://github.com/microsoft/LLMLingua}{github.com/microsoft/LLMLingua} (README call).}
\label{alg:llmlingua2}
\small
\begin{algorithmic}[1]
\Require uncompressed context $S$, rate $r$ (the knob), classifier \texttt{xlm-roberta-large-meetingbank}
\For{each span body $b$ of $S$ (independently)}
  \State split $b$ into chunks of 512 classifier tokens; $p_t \gets$ the classifier's probability that token $t$ is kept
  \State keep the tokens whose $p_t$ is at or above the $\lfloor 100(1-r)+1 \rfloor$-th percentile of $p$, always keeping \texttt{\textbackslash n} and \texttt{?}
  \State $b' \gets$ the kept tokens in order \Comment{indentation and words inside a line may be deleted}
\EndFor
\State \Return the spans with bodies $b'$; if no 1\%-level of $r$ lands within $\pm$2\%, the first $j$ spans take the next level
\end{algorithmic}
\end{algorithm}

\begin{algorithm}[t]
\caption{\textbf{LongLLMLingua} \citep{jiang2024longllmlingua}: question-aware coarse-to-fine token pruning with a causal LM (Qwen2.5-7B-Instruct), budget-matched with Algorithm~\ref{alg:fit} on its target token count. Paper: \href{https://arxiv.org/abs/2310.06839}{arXiv:2310.06839}; code: \href{https://github.com/microsoft/LLMLingua}{github.com/microsoft/LLMLingua} (documented LongLLMLingua settings).}
\label{alg:longllmlingua}
\small
\begin{algorithmic}[1]
\Require span bodies $d_1,\dots,d_n$ of $S$ (the documents), issue $q$ (the question), target $t$ (the knob)
\State score each $d_k$ by the perplexity of $q$ given $d_k$ (\texttt{rank\_method=longllmlingua}, question after the documents); sort the documents by it
\State select documents in that order until the budget $t$ plus 100 tokens is reached
\State give each selected document a token budget that grows with its rank (dynamic ratio 0.3)
\State prune tokens iteratively in segments of 200, removing the tokens with the lowest question-conditioned contrastive perplexity, until $t$ tokens remain
\State \Return each kept token mapped back to its span, documents in relevance order
\end{algorithmic}
\end{algorithm}

\begin{algorithm}[t]
\caption{\textbf{LongCodeZip} \citep{shi2025longcodezip}: function-level then block-level pruning for code with Qwen2.5-Coder-7B-Instruct, budget-matched with Algorithm~\ref{alg:fit} on its rate. Paper: \href{https://doi.org/10.1109/ase63991.2025.00020}{ASE 2025}; code: \href{https://github.com/YerbaPage/LongCodeZip}{github.com/YerbaPage/LongCodeZip} (the paper's code-completion settings).}
\label{alg:longcodezip}
\small
\begin{algorithmic}[1]
\Require span bodies of $S$ joined by the official separator, issue $q$ (the query), rate $r$ (the knob), fine ratio $0.8$
\State split the code into functions; score each by how much it lowers the perplexity of $q$ (conditional perplexity)
\State keep the highest-scoring functions within a budget of $r$ times the input (dynamic ratio 0.2, plus 100 tokens); replace each omitted function with \texttt{\# ...}
\For{each kept function}
  \State split it into blocks by the entropy of a small model (Qwen2.5-Coder-0.5B-Instruct)
  \State choose blocks by a knapsack over their conditional-perplexity importance ($\beta=0.5$), keeping a fine ratio of its tokens
\EndFor
\State \Return the kept code regrouped per file; a file whose functions are all omitted is dropped; if $r = 0.995$ still overshoots, raise the fine ratio toward 1
\end{algorithmic}
\end{algorithm}

\begin{algorithm}[t]
\caption{\textbf{SWE-Pruner} \citep{wang2026swepruner}: a 0.6B line pruner trained for coding agents (Qwen3-Reranker backbone), run in its single-turn mode with the issue as its goal. Paper: \href{https://arxiv.org/abs/2601.16746}{arXiv:2601.16746}; code and model: \href{https://github.com/Ayanami1314/swe-pruner}{github.com/Ayanami1314/swe-pruner}.}
\label{alg:swepruner}
\small
\begin{algorithmic}[1]
\Require uncompressed context $S$ (the compressors' input), issue $q$, budget $B_i$
\State $g \gets$ the first 1,024 tokens of $q$ \Comment{goal hint}
\For{each span $s$ of $S$}
  \State score every line of $s$ given $g$ (8,192-token windows, 50-token overlap; token scores averaged per line)
\EndFor
\State binary-search the threshold $\tau$ over the observed line scores for the rendering closest to $B_i$
\State \Return each span's header, its lines with score $\ge\tau$ verbatim (single-line gaps kept), each pruned run as ``(filtered $n$ lines)''; spans with no kept line dropped
\end{algorithmic}
\end{algorithm}

\begin{algorithm}[t]
\caption{\textbf{Chunk retrieval: BM25 \citep{robertson2009bm25} and dense retrieval with CodeRankEmbed \citep{suresh2025cornstack}.} Papers: \href{https://doi.org/10.1561/1500000019}{BM25}; \href{https://openreview.net/forum?id=iyJOUELYir}{CoRNStack (ICLR 2025)}; model: \href{https://huggingface.co/nomic-ai/CodeRankEmbed}{huggingface.co/nomic-ai/CodeRankEmbed}.}
\label{alg:chunks}
\small
\begin{algorithmic}[1]
\Require issue $q$, repository $R$ at the base commit, budget $B_i$
\For{each source file of $R$ (at most 512 KB)}
  \State chunks $\gets$ for Python, its top-level functions and classes (decorators included) and the code between them; for other files and Python that does not parse, 60-line windows; chunks over 4K tokens are split into 60-line windows
\EndFor
\State \textbf{BM25:} score each chunk (its path and text) against $q$ with $k_1{=}1.2$, $b{=}0.75$ over code terms (identifiers split at case changes and underscores)
\State \textbf{Dense:} score each chunk by the cosine between CodeRankEmbed's embeddings of $q$ (prefixed with its documented query prompt) and of the chunk (4,096-token limit)
\State $m \gets [\,]$; \textbf{for} chunks in descending score (BM25: score $>0$) \textbf{do} add the chunk while $|m| \le B_i$
\State cut the first chunk that does not fit to whole lines if more than 64 tokens remain, then stop
\State \Return $m$, every line verbatim
\end{algorithmic}
\end{algorithm}

\begin{algorithm}[t]
\caption{\textbf{File-level BM25}, SWE-bench's retrieval baseline \citep{jimenez2024swebench}, reproduced from \texttt{swebench.inference.make\_datasets} (\texttt{swebench} 4.0.3). Paper: \href{https://openreview.net/forum?id=VTF8yNQM66}{SWE-bench (ICLR 2024)}; code: \href{https://github.com/SWE-bench/SWE-bench}{github.com/SWE-bench/SWE-bench}.}
\label{alg:bm25files}
\small
\begin{algorithmic}[1]
\Require issue $q$, repository $R$ at the base commit, budget $B_i$
\State documents $\gets$ every Python file whose path does not contain the word \texttt{test}, \texttt{tests} or \texttt{testing}, as its path followed by its contents
\State rank the documents against $q$ with BM25 ($k_1{=}0.9$, $b{=}0.4$, Pyserini's defaults)
\State $m \gets$ the top 20 files added whole in rank order, skipping any file that does not fit in what remains of $B_i$
\State \Return $m$ (it can land under $B_i$)
\end{algorithmic}
\end{algorithm}

\begin{algorithm}[t]
\caption{\textbf{SweRank} \citep{reddy2026swerank}: retrieve-and-rerank function localization (SweRankEmbed-Small, 137M; SweRankLLM-Small, 7B listwise), with the official reranking prompt. Paper: \href{https://arxiv.org/abs/2505.07849}{arXiv:2505.07849}; code: \href{https://github.com/SalesforceAIResearch/SweRank}{github.com/SalesforceAIResearch/SweRank}; models: \href{https://huggingface.co/Salesforce/SweRankEmbed-Small}{SweRankEmbed-Small}, \href{https://huggingface.co/Salesforce/SweRankLLM-Small}{SweRankLLM-Small}.}
\label{alg:swerank}
\small
\begin{algorithmic}[1]
\Require issue $q$, repository $R$ at the base commit, budget $B_i$
\State $U \gets$ every top-level function and class method, decorators included, in non-test Python files of $R$
\State rank $U$ by the cosine between SweRankEmbed's embeddings of $q$ and of each function; $T \gets$ the top 100
\For{windows of 10 from the bottom of $T$ to the top, in steps of 5}
  \State SweRankLLM ranks the window's functions (each cut to 1,024 tokens) by ``relevance to contain the faults causing the GitHub issue''; reorder $T$ (greedy decoding)
\EndFor
\State add whole functions in the order of $T$, then the rest by similarity, while the context fits $B_i$; cut the first that does not fit to whole lines if more than 64 tokens remain, then stop
\State \Return the packed functions, verbatim, each with its path and line range
\end{algorithmic}
\end{algorithm}

\begin{algorithm}[t]
\caption{\textbf{The reference contexts} (bounds, not methods). The gold patch is used only here and in offline measurement; it is never shown to any other context.}
\label{alg:bounds}
\small
\begin{algorithmic}[1]
\Require issue $q$, repository $R$ at the base commit, gold patch $p^\star$
\State \textbf{Gold files (upper bound):} \Return every pre-existing file that $p^\star$ modifies (its \texttt{-{}-{}- a/} paths), in full and verbatim; not fitted to $B_i$
\State \textbf{No source (lower bound):} \Return the empty context; the model sees only $q$
\end{algorithmic}
\end{algorithm}

\section{What Went Wrong and What We Learned}
\label{app:tacit}

\paragraph{Grading with exit codes is not grading resolution.}
Before this study, our resolution numbers came from an executor that marked an issue resolved when the repository's test command exited with code 0, without applying the issue's test patch or checking the FAIL\_TO\_PASS and PASS\_TO\_PASS lists.
On 6 issues we checked, an \emph{empty} patch counted as resolved on 4 under that rule and on 0 under official grading.
That rule largely measures whether the existing test suite passes.
All results in this paper use official grading, and every issue was checked in advance to be resolved by its gold patch and not by an empty patch.

\paragraph{Budget matching is fiddly.}
Rate knobs are coarse (LLMLingua-2 quantizes its threshold to 1\%; LongCodeZip's rate cannot be set to 1.0 because of an upstream bug), so we search a fixed grid with interpolation and record misses rather than re-running with other settings.
Span headers count toward the budget and are kept verbatim for every method, so compressors never lose file identity.

\paragraph{Infrastructure.}
Model-based compressors ran out of memory on the largest contexts; we released GPU memory between issues and retried those issues in a fresh process with expandable memory segments, and LongCodeZip still failed on 4 of Claude's 300 contexts, even on a 48\,GB A6000 (counted as unresolved; in the worst case for \ours{} the LongCodeZip contrast keeps $p = 0.003$). With Claude, one issue's tests hung in two separate grading runs for every context, although they had passed in the eligibility check; it is counted as unresolved for all contexts, which leaves every paired contrast unchanged.
Parallel container builds into a cold shared cache corrupted it, so builds are serialized.

\section{Planned and Post-hoc Analyses}
\label{app:posthoc}

Four hypotheses, their margins and tests come from a written experiment plan that we did not freeze or time-stamp, so we do not claim pre-registration.
H1 (\ours{} vs.\ the three compressors) and H2 (vs.\ the other budget-matched exact-text contexts: the size-matched single draw, chunk BM25 and dense retrieval) use one-sided exact McNemar tests with Holm correction within each family; H3 tests the evidence step; H4 tests non-inferiority to the uncompressed single-draw context (lower 95\% bound of the resolution ratio above 0.85 and upper bound of the token ratio below 0.75).
H1 holds; H2 holds for both retrievers and fails for the size-matched single draw (\S\ref{sec:selector}); H3 is not supported at $\alpha = 0.05$ ($p = 0.068$); H4 passes with GPT-5.6, while with Claude the token ratio's upper bound (0.757) just misses its margin (Appendix~\ref{app:claude}).
The plan listed the edit-format control and a second model as secondary checks.
Added after the main results, and post hoc: file-level BM25, the search/replace runs for LongLLMLingua, LongCodeZip and dense retrieval, the in-tolerance analysis, the failure decomposition, the controlled tests of \S\ref{sec:controlled}, the second patch sample, the Claude gold-file oracle on the last 200 issues and the offline checks in Table~\ref{tab:w2}.
For the 300-issue corruption test we wrote the design, the primary contrast and its decision rule in a dated file before sending any request.
\rev{SWE-Pruner and SweRank were added after the main results, with the main study's issues, budgets, prompt, post-processing and grader, and the estimator's predictions for them were frozen before any of their patches existed (Appendix~\ref{app:newbaselines}). Table~\ref{tab:hyp} adds them to H1 and H2. Adding them changes three adjusted $p$ values, listed in the caption.}

\begin{table*}[t]
\centering
\footnotesize
\setlength{\tabcolsep}{3pt}
\begin{tabular}{lrrrrrr}
\toprule
 & \multicolumn{3}{c}{GPT-5.6} & \multicolumn{3}{c}{Claude Opus 5} \\ \cmidrule(lr){2-4}\cmidrule(lr){5-7}
Contrast & $\Delta$ [95\% CI] & Won/lost & Holm $p$ & $\Delta$ [95\% CI] & Won/lost & Holm $p$ \\
\midrule
\multicolumn{7}{l}{\emph{H1: ours vs.\ the compressors}} \\
\quad LLMLingua-2 & $-$13.7 {\scriptsize[$-$18.3, $-$9.0]} & 48/7 & $<10^{-4}$ & $-$6.3 {\scriptsize[$-$12.3, $-$0.3]} & 53/34 & 0.0265 \\
\quad LongLLMLingua & $-$8.7 {\scriptsize[$-$12.7, $-$4.7]} & 33/7 & $<10^{-4}$ & $-$17.7 {\scriptsize[$-$23.7, $-$11.7]} & 73/20 & $<10^{-4}$ \\
\quad LongCodeZip & $-$6.3 {\scriptsize[$-$10.3, $-$2.3]} & 30/11 & 0.00432$^{a}$ & $-$10.7 {\scriptsize[$-$16.7, $-$4.7]} & 60/28 & 0.00127$^{b}$ \\
\quad SWE-Pruner$^\dagger$ & $-$2.0 {\scriptsize[$-$6.0, +2.0]} & 22/16 & 0.209 & $-$7.3 {\scriptsize[$-$13.0, $-$1.7]} & 49/27 & 0.0154 \\
\multicolumn{7}{l}{\emph{H2: ours vs.\ the other budget-matched exact-text contexts}} \\
\quad Single draw, size-matched & $-$1.0 {\scriptsize[$-$4.3, +2.3]} & 15/12 & 0.701$^{c}$ & +1.7 {\scriptsize[$-$3.0, +6.3]} & 23/28 & 0.799 \\
\quad BM25 (chunks) & $-$14.7 {\scriptsize[$-$19.7, $-$9.7]} & 53/9 & $<10^{-4}$ & $-$12.7 {\scriptsize[$-$18.3, $-$7.0]} & 60/22 & $<10^{-4}$ \\
\quad Dense (chunks) & $-$21.0 {\scriptsize[$-$26.3, $-$16.0]} & 70/7 & $<10^{-4}$ & $-$13.0 {\scriptsize[$-$19.0, $-$7.0]} & 64/25 & $<10^{-4}$ \\
\quad SweRank$^\dagger$ & $-$0.7 {\scriptsize[$-$5.0, +3.7]} & 23/21 & 0.701 & $-$1.7 {\scriptsize[$-$7.0, +3.7]} & 38/33 & 0.635 \\
\multicolumn{7}{l}{\emph{H3: the evidence step (ours vs.\ ours without evidence)}} \\
\quad Ours without evidence & $-$3.0 {\scriptsize[$-$6.7, +0.7]} & 19/10 & 0.068 & \multicolumn{3}{c}{not run} \\
\multicolumn{7}{l}{\emph{H4: non-inferiority to the uncompressed single draw (resolution ratio $>0.85$, token ratio $<0.75$)}} \\
\quad Resolution ratio (lower bound) & 0.942 (0.878) & & passes & 1.085 (0.985) & & fails \\
\quad Token ratio (upper bound) & 0.684 (0.721) & & & 0.733 (0.757) & & \\
\bottomrule
\end{tabular}
\caption{\rev{\textbf{The four planned hypotheses, with the code-specific baselines added to their families.} $\Delta$: row minus ours (points), 20,000-draw paired issue bootstrap; Won/lost: issues only ours resolves / only the row resolves; Holm $p$: one-sided exact McNemar, Holm-adjusted within the family extended by SWE-Pruner (H1) and SweRank (H2); $^\dagger$: added after the main results. H3 and H4 do not involve the new baselines. H3: one-sided exact McNemar, GPT-5.6 only. H4: RR, resolution ratio (ours over the single draw) with its one-sided 95\% lower bound; TR, token ratio with its upper bound. Holm $p$ changed by the added baselines, published values: $^{a}$0.00216; $^{b}$0.000846; $^{c}$0.351.}}
\label{tab:hyp}
\end{table*}

\section{Localization of the Variants}
\label{app:variants}

\begin{table*}[t]
\centering
\footnotesize
\setlength{\tabcolsep}{6pt}
\begin{tabular}{lrrr@{}lrr@{}l}
\toprule
 & & \multicolumn{3}{c}{GPT-5.6} & \multicolumn{3}{c}{Claude Opus 5} \\
\cmidrule(lr){3-5}\cmidrule(lr){6-8}
Context & Tokens & Res. & \multicolumn{2}{c}{$\Delta$} & Res. & \multicolumn{2}{c}{$\Delta$} \\
\midrule
Ours (CQ2-CARE) & 13.8K & 43.7 & \multicolumn{2}{c}{reference} & 46.7 & \multicolumn{2}{c}{reference} \\
Single draw (Q50-CARE) & 20.1K & 46.3 & +2.7 &  & 43.0 & $-$3.7 &  \\
Single draw, size-matched & 13.3K & 42.7 & $-$1.0 &  & 48.3 & +1.7 &  \\
Union of two draws & 16.1K & 47.7 & +4.0 & $^{*}$ & \multicolumn{3}{c}{not run} \\
Ours without evidence & 10.4K & 40.7 & $-$3.0 &  & \multicolumn{3}{c}{not run} \\
\bottomrule
\end{tabular}
\caption{\textbf{Variants of exact-file selection} (resolved \%, 300 issues). The two single-draw variants and the union keep whole files from other localization sets, all with the same evidence step; ``ours without evidence'' drops that step. Tokens: GPT-5.6 contexts. $\Delta$: row minus ours (points); $^{*}p<0.05$ in two-sided exact McNemar tests. The union and the variant without evidence were run with GPT-5.6 only.}
\label{tab:variants}
\end{table*}
\begin{table}[t]
\centering
\footnotesize
\setlength{\tabcolsep}{2.2pt}
\begin{tabular}{lrrrr}
\toprule
Localized set & Files & Precision & Recall & F1 \\
\midrule
Single draw (Q50) & 3.40 & 0.367 & 0.650 & 0.430 \\
Single draw, size-matched & 1.69 & 0.696 & 0.608 & 0.600 \\
Two-draw agreement (ours) & 1.71 & 0.705 & 0.631 & 0.613 \\
Union of two draws & 2.23 & 0.655 & 0.674 & 0.596 \\
1-of-5 agreement & 2.53 & 0.627 & 0.689 & 0.584 \\
2-of-5 agreement & 2.15 & 0.673 & 0.680 & 0.610 \\
3-of-5 agreement & 1.89 & 0.698 & 0.658 & 0.617 \\
4-of-5 agreement & 1.71 & 0.717 & 0.638 & 0.623 \\
5-of-5 agreement & 1.59 & 0.716 & 0.604 & 0.607 \\
\midrule
SweRank$^\dagger$, top file & 1.00 & 0.770 & 0.526 & 0.594 \\
SweRank$^\dagger$, top 2 files & 1.99 & 0.477 & 0.616 & 0.507 \\
SweRank$^\dagger$, all files shown & 11.76 & 0.163 & 0.717 & 0.239 \\
\bottomrule
\end{tabular}
\caption{Changed-file localization against the pre-existing files the gold patch modifies (mean 1.86 per issue), 300 issues, GPT-5.6. Two-draw agreement beats a single draw by +0.184 F1 [+0.161, +0.206], but a single draw cut to the same size is within 0.013 [-0.005, +0.031]. SweRank's files are taken in the order its ranked functions first reach them; SWE-Pruner prunes the single draw's context and keeps its localization. $^\dagger$: added after the main results.}
\label{tab:loc}
\end{table}

Table~\ref{tab:loc} gives changed-file localization for every localized set of \S\ref{sec:selector}, including the $t$-of-5 agreement thresholds of Figure~\ref{fig:budgets}.
Against the uncompressed single draw, \ours{} uses 0.68$\times$ the tokens with GPT-5.6 (upper bound 0.72) and is non-inferior (resolution ratio 0.94, lower bound 0.878 above the planned margin of 0.85).

\section{Details of the Second-Model Run}
\label{app:claude}

Claude's patches on LLMLingua-2 contexts apply 80.7\% of the time (GPT-5.6: 70.7\%), while its patches apply less often in general (83.3\% on our contexts, against 98.0\% for GPT-5.6), so applicability separates the contexts less; its hunks carry slightly less unchanged context (median 5 vs.\ 6 lines), but we have not isolated the cause.
Heavy deletion has a different failure with Claude: 36 of its 300 answers on LongLLMLingua contexts failed to parse, 31 of them placeholders or empty patches such as \texttt{"..."}, \texttt{"<unified diff>"} or \texttt{"d i ff . . ."} that echo the compressed text; 35 of these 36 issues received well-formed patches on LLMLingua-2 contexts.
Claude's localization keeps more files, so its contexts are about twice as large as GPT's (27.1K vs.\ 13.8K mean tokens for \ours{}). Against the uncompressed single draw, \ours{} resolves more (46.7\% vs.\ 43.0\%) at 0.73$\times$ the tokens, but the token ratio's upper bound (0.757) just misses H4's 0.75 margin. With Claude, the gold-file oracle resolves 46.3\%, about as often as \ours{} (46.7\%); its patches apply on 83.0\% of issues. We ran it on the last 200 issues after the main analysis, with the same code and settings, because the first run's budget had stopped before it. Table~\ref{tab:claude} gives the full comparison.

\section{Details of the Analyses}
\label{app:analysis}

\begin{table}[t]
\centering
\footnotesize
\setlength{\tabcolsep}{2.4pt}
\begin{tabular}{lrrrr@{}l}
\toprule
Context & $\varepsilon$ & App. & Res. & \multicolumn{2}{c}{$\Delta$} \\
\midrule
\multicolumn{6}{l}{\textbf{(a) All 300 issues}} \\
Ours (CQ2-CARE) & 100.0 & 98.0 & 43.7 & \multicolumn{2}{c}{reference} \\
\quad second patch sample & 100.0 & 98.0 & 43.7 & 0.0 &  \\
\quad all lines corrupted & 0.3 & 78.7 & 34.0 & $-$9.7 & $^{***}$ \\
\multicolumn{6}{l}{\emph{From our candidate files:}} \\
\quad LongCodeZip & 98.3 & 95.7 & 44.7 & +1.0 &  \\
\quad LLMLingua-2, lines & 100.0 & 89.7 & 40.7 & $-$3.0 &  \\
\quad LLMLingua-2, tokens & 41.2 & 80.0 & 36.3 & $-$7.3 & $^{**}$ \\
\multicolumn{6}{l}{\emph{From a single draw (main study):}} \\
\quad LongCodeZip & 96.8 & 89.7 & 37.3 & $-$6.3 & $^{**}$ \\
\quad LLMLingua-2, tokens & 17.5 & 70.7 & 30.0 & $-$13.7 & $^{***}$ \\
\midrule
\multicolumn{6}{l}{\textbf{(b) 100-issue subset}} \\
Ours (CQ2-CARE) & 100.0 & 98.0 & 41.0 & \multicolumn{2}{c}{reference} \\
\quad second patch sample & 100.0 & 98.0 & 44.0 & +3.0 &  \\
\quad 25\% lines corrupted & 76.1 & 85.0 & 39.0 & $-$2.0 &  \\
\quad 50\% lines corrupted & 51.5 & 83.0 & 35.0 & $-$6.0 &  \\
\quad all lines corrupted & 0.4 & 80.0 & 38.0 & $-$3.0 &  \\
LLMLingua-2, tokens & 23.2 & 71.0 & 28.0 & \multicolumn{2}{c}{reference} \\
\quad whole lines & 100.0 & 86.0 & 36.0 & +8.0 & $^{**}$ \\
\bottomrule
\end{tabular}
\caption{\textbf{Controlled tests of verbatim text and gold-file recall} (GPT-5.6, post hoc). Corrupted rows: our own files with a share of lines each losing 20\% of their tokens. Candidate files: the union of our two localization draws with the evidence step, compressed to $B_i$. Whole lines: LLMLingua-2's own token scores averaged per line, whole verbatim lines kept up to $B_i$. $\Delta$: resolution in points against the indented block's reference row (in (b), the whole-line row against token-level LLMLingua-2); $\varepsilon$ (verbatim lines), App.\ (patches apply) and Res.\ (resolved) in \%; $^{***}p<0.001$, $^{**}p<0.01$, $^{*}p<0.05$ in two-sided exact McNemar tests.}
\label{tab:controlled}
\end{table}

\paragraph{Controlled tests.}
\rev{Table~\ref{tab:controlled} gives both tests in full.} All figures are GPT-5.6 on 300 issues, with 95\% issue-bootstrap intervals. Corrupting our contexts lowers resolution by 9.7 points [5.7, 14.0]; the second patch sample differs by 0.0 [$-$3.3, 3.3]. Our candidate files, the union of our two localization draws with the evidence step, average 16.1K tokens, against 20.1K for the single-draw context. Cutting them to $B_i$, LongCodeZip differs from \ours{} by +1.0 points [$-$2.7, 4.7], and LLMLingua-2 by $-$3.0 [$-$6.3, 0.3] with whole lines and $-$7.3 [$-$11.7, $-$3.0] with tokens. On the 177 issues where the union exceeds $B_i$, so that the compressor must cut, token-level LLMLingua-2 trails by 13.0 points, while LongCodeZip is 0.6 points ahead. Gold-file recall falls from 78.3\% to 72.2\% when LongCodeZip cuts the single-draw context, but only from 79.9\% to 78.3\% when it cuts our candidates. On the 100-issue subset, whole lines chosen by LLMLingua-2's scores from the single-draw context raise resolution from 28.0\% to 36.0\% ($p = 0.008$).

\paragraph{Robustness.}
The compressors miss the $\pm$2\% window on 26.3--34.3\% of issues; within it, LongCodeZip trails by 6.6 points on 197 issues (Holm $p = 0.010$) and LLMLingua-2 by 16.7 on 221. With search/replace edits on the 100-issue subset, the baselines trail by 9.0 (LongCodeZip) to 19.0 points (chunk BM25), and resolution moved by at most 4.0 points from the diff format (LongLLMLingua); a quote matched no unique location in the real file on 45 issues for LLMLingua-2, 32 for LongLLMLingua, 16 for LongCodeZip and 4 for \ours{}. In the rate sweep, LongCodeZip reaches at most 37.0\% (at 12.6K tokens), LLMLingua-2 32.0\% and chunk BM25 25.0\% (at about 13K), while exact-file contexts resolve 41.0--44.0\% between 11.9K and 15.1K tokens. The second patch sample flips 8.7--10.0\% of outcomes; averaged over both samples, \ours{} leads LongCodeZip by 6.3 points [3.0, 9.8] and LLMLingua-2 by 14.7 [10.7, 18.8] (sign tests, $p \le 0.002$).

\section{\rev{The Trained Methods and the Estimator}}
\label{app:newbaselines}

\paragraph{\rev{Overview.}}
\label{sec:trained}
\begin{table}[t]
\centering
\footnotesize
\setlength{\tabcolsep}{5pt}
\begin{tabular}{@{}llrrr@{}}
\toprule
Context & Model & Obs. & Pred. & 95\% PI \\
\midrule
\multicolumn{5}{@{}l}{\emph{Predicted before any of their patches}} \\
SWE-Pruner & GPT-5.6 & 41.7 & 40.4 & 35.7--45.0 \\
 & Claude & 39.3 & 42.0 & 36.0--47.7 \\
SweRank & GPT-5.6 & 43.0$^{\times}$ & 37.7 & 33.0--42.0 \\
 & Claude & 45.0$^{\times}$ & 37.5 & 31.7--43.0 \\
\midrule
\multicolumn{5}{@{}l}{\emph{Post hoc: $\kappa$ counted inside functions}} \\
SweRank & GPT-5.6 & 43.0 & 39.7 & 35.3--44.0 \\
 & Claude & 45.0 & 39.5 & 33.7--45.0 \\
\bottomrule
\end{tabular}
\caption{\textbf{The estimator's predictions for two methods it has not seen.} Observed and predicted resolution (\%) with the 95\% prediction interval (PI); $^{\times}$: observed outside the interval. The main contexts, each left out of the fit, land within their intervals for 11 of 11 GPT-5.6 and 8 of 9 Claude contexts (Table~\ref{tab:est-heldout}).}
\label{tab:est}
\end{table}

\begin{figure*}[t]\centering\figsafe{width=\textwidth}{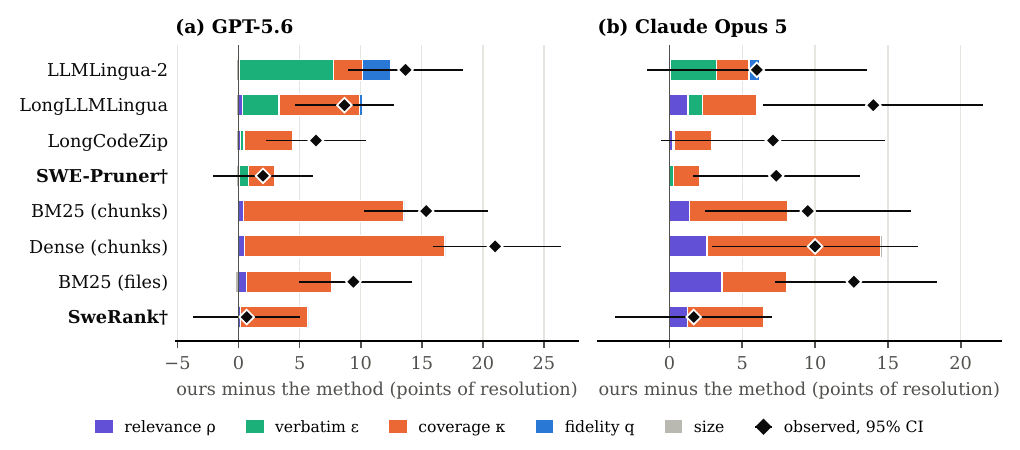}
\caption{\rev{\textbf{What the predicted lead of \ours{} over each method is made of}, split exactly among the estimator's variables (Shapley values), with the observed gap and its 95\% interval.}}\label{fig:shap}\end{figure*}

\rev{To test whether gold-file recall and verbatim text account for resolution, we fit an estimator.
For each model, a logistic regression over (issue, context) pairs predicts resolution from a context's $\rho$, $\varepsilon$ and size, the share $\kappa$ of the gold patch's changed lines it shows, and whether the issue is resolved with no source and with the gold files.
Each main context, left out of the fit, lands within its 95\% prediction interval for 11 of 11 GPT-5.6 and 8 of 9 Claude contexts, and the estimator attributes the retrievers' deficit mainly to changed lines they do not show and the compressors' to altered and missing lines (Figure~\ref{fig:shap}).}

\rev{Its predictions for the two trained methods, frozen before any of their patches existed, test it on methods it has not seen (Table~\ref{tab:est}).
SWE-Pruner lands within its intervals, but SweRank resolves 5.3 and 7.5 points more than predicted, apparently because the changed lines it leaves out, imports, module constants and class attributes outside functions, cost less than the estimator charges.
Counting $\kappa$ only on changed lines inside functions moves SweRank within its intervals without refitting, but we chose this change after seeing the miss, so it still needs a test on a method the estimator has not seen.}

\paragraph{\rev{Within the budget window.}}
\rev{SWE-Pruner lands within $\pm$2\% of $B_i$ on 256 of 300 GPT-5.6 issues and SweRank on 289. On the rest, no pruning threshold or next whole function fits.
On the issues within the window, their gaps to \ours{} are 1.6 and 1.0 points with GPT-5.6 and 7.0 and 1.1 with Claude (Table~\ref{tab:robust}), about the same as on all issues, so the small budget shortfalls do not drive the result.}

\paragraph{\rev{Issues that only one of them resolves.}}
\rev{With GPT-5.6, \ours{} resolves 23 issues that SweRank does not, and SweRank resolves 21 that \ours{} does not.
With Claude the counts are 38 and 33.
Against SWE-Pruner they are 22 and 16 with GPT-5.6, and 49 and 27 with Claude.
Table~\ref{tab:goldfit} splits the issues by whether the gold files fit in $B_i$, for both models.}

\paragraph{\rev{The estimator's variables.}}
\rev{Let $\Lambda^\star$ be the changed lines of the gold patch: for each pre-existing source file it modifies (documentation and changelogs excluded), the lines the patch removes, and for a pure insertion the line before it.
A changed line $\ell$ is \emph{shown} in a context $m$ when a span of $m$ covers its line number and the span's text contains $\ell$ verbatim, or a line that keeps at least three quarters of $\ell$'s identifiers and literals.
We write $\ell'$ for that line of $m$ and $\mathrm{tok}(\cdot)$ for the multiset of identifiers, literals and punctuation of a line. Then
\begin{align}
  \kappa(m) &= \frac{|\{\ell \in \Lambda^\star : \ell \text{ shown in } m\}|}{|\Lambda^\star|}, \label{eq:kappa}\\
  q(m) &= \operatorname*{mean}_{\ell \text{ shown}} \frac{|\mathrm{tok}(\ell) \cap \mathrm{tok}(\ell')|}{|\mathrm{tok}(\ell)|}. \label{eq:q}
\end{align}
So $\kappa$ is the share of the changed lines that the context shows, and $q$ is the share of each shown changed line's tokens that it keeps, with $q = 1$ when every shown changed line is verbatim and $q = 0$ when none is shown.
$\kappa_{\mathrm{fn}}$ is \eqref{eq:kappa} counted only over the changed lines inside a function or method body at the base commit.
Like $\rho$ and $\varepsilon$, these use the gold patch, so they are computed after the fact and never shown to a model.}

\begin{figure*}[t]\centering\figsafe{width=\textwidth}{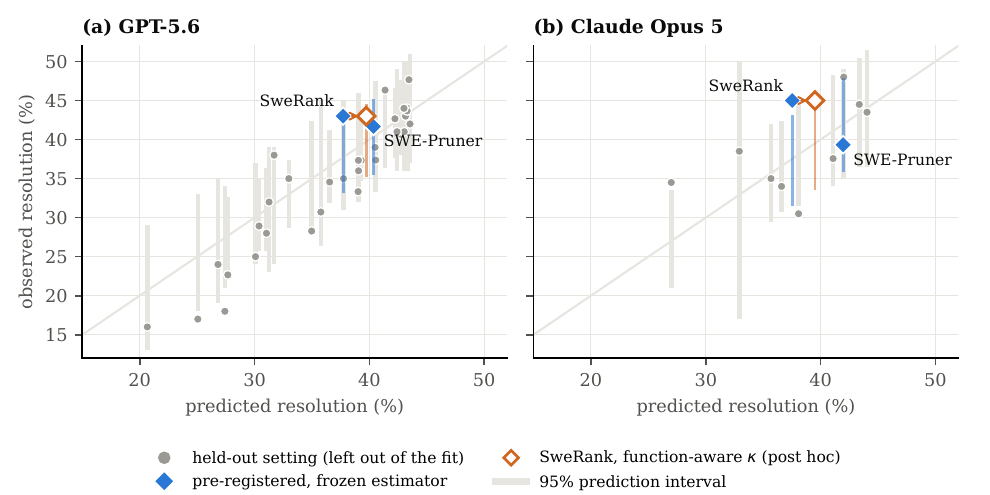}
\caption{\rev{\textbf{Predicted against observed resolution for every setting left out of the fit}, with 95\% prediction intervals. Blue: the four frozen predictions. Orange: SweRank with coverage counted inside functions.}}\label{fig:calib}\end{figure*}

\paragraph{\rev{The model and its accuracy.}}
\rev{For each model, we fit one logistic regression over the (issue, context) pairs of its main contexts. It predicts resolution from $y_\emptyset$ and $y_\star$, whether the issue is resolved with no source and with the gold files, and from $\rho$, $\varepsilon$, $\kappa$, $q$ and $\log|m| - \mu$, the log token count centred on the fit's mean (ridge $10^{-3}$, Table~\ref{tab:est-coef}).
The predicted resolution of a method is the mean of its per-issue probabilities.
Its 95\% prediction interval is the range that covers 95\% of the rates obtained by refitting on 20,000 bootstrap resamples of the issues and drawing one patch outcome per issue, so it covers both the uncertainty of the fit and the noise of a single patch sample.
Each main context, left out of the fit and predicted on its own issues, falls inside its interval for 11 of 11 GPT-5.6 contexts and 8 of 9 Claude contexts, with RMSE 3.3 and 5.1 points (Tables~\ref{tab:est-heldout} and~\ref{tab:est-summary}, Figure~\ref{fig:calib}).
The Claude miss is LongLLMLingua, whose placeholder patches (Appendix~\ref{app:claude}) the variables do not see.
Over all 33 GPT-5.6 settings left out, including the 100-issue rate sweep, the second patch samples and the controlled text changes, the RMSE is 3.9 points and 31 observations fall inside their intervals.}

\begin{table}[t]
\centering
\scriptsize
\setlength{\tabcolsep}{3pt}
\begin{tabular}{lrr}
\toprule
Term & GPT-5.6 & Claude Opus 5 \\
\midrule
intercept & $-$5.74 {\scriptsize[$-$6.73, $-$5.01]} & $-$3.80 {\scriptsize[$-$4.75, $-$3.07]} \\
no source resolves, $y_\emptyset$ & +1.34 {\scriptsize[+0.62, +2.30]} & +1.29 {\scriptsize[+0.75, +1.83]} \\
gold files resolve, $y_\star$ & +3.59 {\scriptsize[+3.12, +4.20]} & +2.10 {\scriptsize[+1.72, +2.57]} \\
gold-file recall, $\rho$ & +0.19 {\scriptsize[$-$0.54, +0.89]} & +0.88 {\scriptsize[+0.28, +1.52]} \\
verbatim lines, $\varepsilon$ & +0.81 {\scriptsize[+0.41, +1.25]} & +0.20 {\scriptsize[$-$0.19, +0.61]} \\
changed lines shown, $\kappa$ & +1.86 {\scriptsize[+1.08, +2.86]} & +1.26 {\scriptsize[+0.56, +2.07]} \\
their token retention, $q$ & +0.88 {\scriptsize[$-$0.01, +1.70]} & +0.18 {\scriptsize[$-$0.59, +0.94]} \\
size, $\log|m| - \mu$ & $-$0.29 {\scriptsize[$-$0.53, $-$0.06]} & +0.11 {\scriptsize[$-$0.16, +0.40]} \\
\bottomrule
\end{tabular}
\caption{\rev{\textbf{The estimator's coefficients} (log-odds per unit of each variable, with the 95\% interval of a 20,000-draw issue bootstrap), one fit per model over the (issue, context) pairs of its main contexts.}}
\label{tab:est-coef}
\end{table}
\begin{table}[t]
\centering
\scriptsize
\setlength{\tabcolsep}{2pt}
\begin{tabular}{lrrrr}
\toprule
 & \multicolumn{2}{c}{GPT-5.6} & \multicolumn{2}{c}{Claude Opus 5} \\ \cmidrule(lr){2-3}\cmidrule(lr){4-5}
Context & Obs. & Pred.\ [95\% PI] & Obs. & Pred.\ [95\% PI] \\
\midrule
Ours & 43.7 & 43.2 {\scriptsize\color{gray}[38.7, 48.0]} & 46.7 & 44.0 {\scriptsize\color{gray}[38.0, 50.0]} \\
Ours without evidence & 40.7 & 42.8 {\scriptsize\color{gray}[38.0, 47.7]} & \multicolumn{2}{c}{--} \\
Union of two draws & 47.7 & 43.4 {\scriptsize\color{gray}[38.7, 48.3]} & \multicolumn{2}{c}{--} \\
Single draw (Q50) & 46.3 & 41.4 {\scriptsize\color{gray}[36.3, 46.3]} & 43.0 & 44.9 {\scriptsize\color{gray}[38.3, 51.3]} \\
Size-matched draw & 42.7 & 42.2 {\scriptsize\color{gray}[37.7, 46.7]} & 48.3 & 42.7 {\scriptsize\color{gray}[36.7, 48.7]} \\
\midrule
LLMLingua-2 & 30.0 & 35.5 {\scriptsize\color{gray}[26.0, 44.3]} & 40.3 & 33.9 {\scriptsize\color{gray}[18.0, 50.7]} \\
LongLLMLingua & 35.0 & 33.0 {\scriptsize\color{gray}[28.7, 37.3]} & 29.0 & 38.4$^{\times}$ {\scriptsize\color{gray}[32.7, 44.0]} \\
LongCodeZip & 37.3 & 39.3 {\scriptsize\color{gray}[34.7, 43.7]} & 36.5 & 42.1 {\scriptsize\color{gray}[36.1, 48.0]} \\
SWE-Pruner$^\dagger$ & 41.7 & 40.4 {\scriptsize\color{gray}[35.7, 45.0]} & 39.3 & 42.0 {\scriptsize\color{gray}[36.0, 47.7]} \\
\midrule
BM25 (chunks) & 28.9 & 30.4 {\scriptsize\color{gray}[25.7, 35.0]} & 34.0 & 35.8 {\scriptsize\color{gray}[30.3, 41.3]} \\
Dense (chunks) & 22.7 & 27.7 {\scriptsize\color{gray}[22.7, 32.7]} & 33.7 & 28.7 {\scriptsize\color{gray}[23.0, 34.3]} \\
BM25 (files) & 34.6 & 36.5 {\scriptsize\color{gray}[31.9, 41.6]} & 34.0 & 36.6 {\scriptsize\color{gray}[30.7, 42.3]} \\
SweRank$^\dagger$ & 43.0 & 37.7$^{\times}$ {\scriptsize\color{gray}[33.0, 42.0]} & 45.0 & 37.5$^{\times}$ {\scriptsize\color{gray}[31.7, 43.0]} \\
\bottomrule
\end{tabular}
\caption{\rev{\textbf{Observed and predicted resolution (\%) of every context.} Each main context is left out of the fit and predicted on its own issues (GPT-5.6: RMSE 3.3 points, 11/11 inside the interval; Claude: 5.1, 8/9); the two added methods ($^\dagger$) are predicted from the full fit, frozen before any of their patches existed. PI: 95\% prediction interval, 20,000 bootstrap refits with one simulated patch per issue. $^{\times}$: observed outside it.}}
\label{tab:est-heldout}
\end{table}
\begin{table}[t]
\centering
\footnotesize
\setlength{\tabcolsep}{3.5pt}
\begin{tabular}{lrrrr}
\toprule
Settings left out of the fit & $n$ & In 95\% PI & RMSE & Bias \\
\midrule
GPT, contexts left out & 11 & 11/11 & 3.3 & $-$0.5 \\
GPT, other budgets & 14 & 12/14 & 4.7 & $-$3.5 \\
GPT, new patch samples & 4 & 4/4 & 1.8 & $-$0.8 \\
GPT, altered text & 4 & 4/4 & 3.8 & $-$0.3 \\
Claude, contexts left out & 9 & 8/9 & 5.1 & $-$0.2 \\
Frozen (both) & 4 & 2/4 & 4.8 & +2.9 \\
\bottomrule
\end{tabular}
\caption{\rev{\textbf{Accuracy by kind of setting left out of the fit} (points. Bias: mean of observed minus predicted). Over all 33 GPT-5.6 settings the RMSE is 3.9 points.}}
\label{tab:est-summary}
\end{table}

\paragraph{\rev{The frozen predictions.}}
\rev{We built the contexts of SWE-Pruner and SweRank, computed the four predictions and froze them (file hash and time stamp in the repository) before generating any of their patches.
The estimator predicts both SWE-Pruner rates inside their intervals and underpredicts SweRank by 5.3 and 7.5 points (Table~\ref{tab:est-prereg}).
The misses are concentrated where SweRank shows the whole function to edit but leaves out changed lines outside functions, such as imports, module constants and class attributes.
Counting $\kappa$ over the changed lines inside functions only, without refitting, moves both SweRank predictions to within their intervals and leaves the fit to the main study and its accuracy on the left-out contexts unchanged.
This correction was chosen after the outcomes, so it must be tested on a method the estimator has not seen.}

\begin{figure*}[t]\centering\figsafe{width=\textwidth}{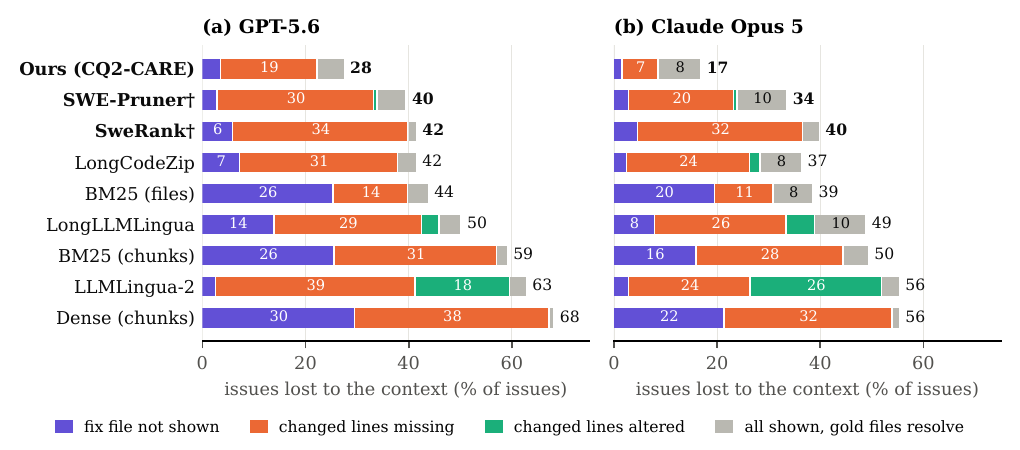}
\caption{\rev{\textbf{Where each context loses the fix.} Unresolved issues attributed to one case: no file with a changed line shown, a changed line missing, a changed line altered, or everything shown while the gold files resolve the issue (\% of issues). $^\dagger$: added after the main results.}}\label{fig:causes}\end{figure*}

\paragraph{\rev{Where each context leaves the fix.}}
\rev{Table~\ref{tab:cmp} puts \ours{} and the two trained methods side by side in the estimator's variables, and sorts every issue into one of four cases by what the context shows of the gold patch's changed lines: no file that contains one, some of them missing, all of them shown but altered, or all of them shown verbatim.
Once every changed line is shown verbatim, the three resolve alike.
Figure~\ref{fig:causes} attributes every unresolved issue of every context to one of these cases, and Figure~\ref{fig:shap} splits the estimator's predicted gap between \ours{} and each method among its variables: for the retrievers the gap is missing changed lines, for the compressors it is altered text and missing lines, and for SweRank it is changed lines outside functions, the part that counting inside functions removes.}

\begin{table*}[t]
\centering
\footnotesize
\setlength{\tabcolsep}{5pt}
\begin{tabular}{lrrr}
\toprule
Setting & Obs. & \begin{tabular}[b]{@{}c@{}}Predicted, frozen\\($\kappa$ on all changed lines)\end{tabular} & \begin{tabular}[b]{@{}c@{}}Predicted, revised\\($\kappa_{\mathrm{fn}}$, inside functions)\end{tabular} \\
\midrule
\multicolumn{4}{l}{\emph{GPT-5.6}} \\
\quad SWE-Pruner$^\dagger$ & 41.7 & 40.4 {\scriptsize\color{gray}[35.7, 45.0]} & 40.8 {\scriptsize\color{gray}[36.3, 45.3]} \\
\quad SweRank$^\dagger$ & 43.0 & 37.7$^{\times}$ {\scriptsize\color{gray}[33.0, 42.0]} & 39.7 {\scriptsize\color{gray}[35.3, 44.0]} \\
\quad \emph{fit to the main study} & & {\color{gray}loss 0.372, AUC 0.901, 3.3 pts, 11/11} & {\color{gray}loss 0.370, AUC 0.902, 3.2 pts, 11/11} \\
\multicolumn{4}{l}{\emph{Claude Opus 5}} \\
\quad SWE-Pruner$^\dagger$ & 39.3 & 42.0 {\scriptsize\color{gray}[36.0, 47.7]} & 42.3 {\scriptsize\color{gray}[36.3, 48.0]} \\
\quad SweRank$^\dagger$ & 45.0 & 37.5$^{\times}$ {\scriptsize\color{gray}[31.7, 43.0]} & 39.5 {\scriptsize\color{gray}[33.7, 45.0]} \\
\quad \emph{fit to the main study} & & {\color{gray}loss 0.516, AUC 0.799, 5.1 pts, 8/9} & {\color{gray}loss 0.519, AUC 0.798, 5.3 pts, 8/9} \\
\bottomrule
\end{tabular}
\caption{\rev{\textbf{The four frozen predictions, and the post-hoc revision that explains the two misses.} Predicted resolution (\%) [95\% prediction interval] of the two added methods, frozen before any of their patches existed, and recomputed with coverage counted only on changed lines inside a function or method ($\kappa_{\mathrm{fn}}$). $^{\times}$: observed outside the interval. The grey rows give each fit's 10-fold cross-validated log loss and AUC on the main study's pairs, and the RMSE (points) and coverage of its leave-one-context-out predictions: the revision changes neither.}}
\label{tab:est-prereg}
\end{table*}
\begin{table}[t]
\centering
\footnotesize
\setlength{\tabcolsep}{2.2pt}
\begin{tabular}{lrrrr}
\toprule
Issues & $n$ & Ours & SWE-Pruner$^\dagger$ & SweRank$^\dagger$ \\
\midrule
\multicolumn{5}{l}{\emph{GPT-5.6}} \\
\quad gold files fit in $B_i$ & 195 & 45.1 & $-$4.1 & $-$1.5 \\
 & & & {\scriptsize\color{gray}[$-$9.2, +1.0]} & {\scriptsize\color{gray}[$-$7.2, +4.1]}\\[1pt]
\quad do not fit & 83 & 43.4 & +2.4 & +1.2 \\
 & & & {\scriptsize\color{gray}[$-$4.8, +9.6]} & {\scriptsize\color{gray}[$-$6.0, +8.4]}\\[1pt]
\multicolumn{5}{l}{\emph{Claude Opus 5}} \\
\quad gold files fit in $B_i$ & 274 & 46.7 & $-$7.7 & $-$0.7 \\
 & & & {\scriptsize\color{gray}[$-$13.5, $-$1.8]} & {\scriptsize\color{gray}[$-$6.6, +5.1]}\\[1pt]
\quad do not fit & 24 & 45.8 & 0.0 & $-$8.3 \\
 & & & {\scriptsize\color{gray}[$-$20.8, +20.8]} & {\scriptsize\color{gray}[$-$25.0, +8.3]}\\[1pt]
\bottomrule
\end{tabular}
\caption{\rev{\textbf{Whether the gold files fit in the budget does not separate the methods.} Issues split by whether the gold files fit in our context's budget $B_i$ for that model. Ours: resolved \%. Other columns: row minus ours (points) with the 95\% paired bootstrap interval. $^\dagger$: added after the main results.}}
\label{tab:goldfit}
\end{table}
\begin{table*}[t]
\centering
\footnotesize
\setlength{\tabcolsep}{4pt}
\begin{tabular}{lrrrrrrrrrrr}
\toprule
 & & & & & & & & \multicolumn{4}{c}{State of the fix: \% of issues / \% of them resolved} \\ \cmidrule(lr){9-12}
Context & Res. & App. & $\rho$ & $\varepsilon$ & $\kappa$ & $\kappa_{\mathrm{fn}}$ & $q$ & \begin{tabular}[b]{@{}c@{}}Relevance\\miss\end{tabular} & \begin{tabular}[b]{@{}c@{}}Coverage\\miss\end{tabular} & \begin{tabular}[b]{@{}c@{}}Fidelity\\loss\end{tabular} & Complete \\
\midrule
\multicolumn{11}{l}{\emph{GPT-5.6}} \\
\quad Ours & 43.7 & 98.0 & 79.0 & 100.0 & 0.83 & 0.83 & 1.00 & 4/0 & 26/28 & 0/-- & 70/52 \\
\quad SweRank$^\dagger$ & 43.0 & 97.3 & 71.7 & 100.0 & 0.63 & 0.73 & 1.00 & 7/18 & 54/37 & 0/-- & 39/56 \\
\quad SWE-Pruner$^\dagger$ & 41.7 & 93.0 & 77.4 & 91.7 & 0.75 & 0.77 & 0.99 & 3/10 & 45/33 & 1/33 & 51/52 \\
\quad LongCodeZip & 37.3 & 89.7 & 72.2 & 96.8 & 0.70 & 0.71 & 0.99 & -- & -- & -- & -- \\
\multicolumn{11}{l}{\emph{Claude Opus 5}} \\
\quad Ours & 46.7 & 83.3 & 84.0 & 100.0 & 0.92 & 0.92 & 1.00 & 2/17 & 14/49 & 0/-- & 84/47 \\
\quad SweRank$^\dagger$ & 45.0 & 89.3 & 75.0 & 100.0 & 0.67 & 0.77 & 1.00 & 5/0 & 54/40 & 0/-- & 42/56 \\
\quad SWE-Pruner$^\dagger$ & 39.3 & 73.7 & 83.9 & 92.1 & 0.84 & 0.85 & 0.99 & 3/0 & 34/40 & 2/67 & 61/40 \\
\quad LongCodeZip & 36.5 & 74.3 & 82.9 & 96.9 & 0.81 & 0.81 & 1.00 & -- & -- & -- & -- \\
\bottomrule
\end{tabular}
\caption{\rev{\textbf{Ours and the two added methods in the estimator's variables} (means over the 300 issues, with LongCodeZip, the strongest original compressor, for reference). Res., App.: resolved and patches that apply (\%). $\rho$, $\varepsilon$, $\kappa$, $q$ as in Appendix~\ref{app:newbaselines}. $\kappa_{\mathrm{fn}}$: coverage of the changed lines inside functions. Each issue is in one state. Relevance miss: no file with a changed line shown. Coverage miss: such a file shown but a changed line missing ($\kappa<1$). Fidelity loss: every changed line shown but altered ($q<1$). Complete: every changed line shown verbatim. $^\dagger$: added after the main results.}}
\label{tab:cmp}
\end{table*}

\section{Additional Tables}
\label{app:tables}

\begin{table*}[t]
\centering
\footnotesize
\setlength{\tabcolsep}{1.6pt}
\begin{tabular}{lrrrrrrrrr}
\toprule
Context & Resolved [95\% CI] & Parse & Applies & Res.$|$app. & Mean tok. & Med.\ tok. & In $\pm$2\% & \$/issue & Build s \\
\midrule
Ours (CQ2-CARE) & 43.7 [38.2, 49.3] & 100.0 & 98.0 & 44.6 & 13,778 & 9,377 & not fitted & 0.151 & 0 \\
Single draw (Q50-CARE) & 46.3 [40.8, 52.0] & 100.0 & 98.0 & 47.3 & 20,143 & 15,168 & not fitted & 0.199 & 0 \\
Single draw, size-matched & 42.7 [37.2, 48.3] & 100.0 & 98.0 & 43.5 & 13,262 & 9,275 & not fitted & 0.121 & 0 \\
Ours without evidence & 40.7 [35.3, 46.3] & 100.0 & 97.7 & 41.6 & 10,367 & 5,756 & not fitted & 0.103 & 0 \\
Union of two draws & 47.7 [42.1, 53.3] & 100.0 & 97.7 & 48.8 & 16,123 & 10,705 & not fitted & 0.153 & 0 \\
LLMLingua-2 & 30.0 [25.1, 35.4] & 99.7 & 70.7 & 42.5 & 13,070 & 8,730 & 73.7 & 0.177 & 13 CPU \\
LongLLMLingua & 35.0 [29.8, 40.6] & 100.0 & 84.0 & 41.7 & 13,154 & 8,788 & 66.0 & 0.176 & 37 GPU \\
LongCodeZip & 37.3 [32.1, 42.9] & 99.3 & 89.7 & 41.6 & 13,126 & 8,930 & 65.7 & 0.168 & 34 GPU \\
BM25 (chunks) & 29.0 [24.2, 34.4] & 100.0 & 86.7 & 33.5 & 13,610 & 9,350 & 98.0 & 0.099 & 0 \\
Dense (chunks) & 22.7 [18.3, 27.7] & 99.7 & 78.3 & 28.9 & 13,659 & 9,354 & 99.0 & 0.106 & 15 GPU \\
SWE-Pruner$^\dagger$ & 41.7 [36.2, 47.3] & 100.0 & 93.0 & 44.8 & 13,361 & 8,957 & 85.3 & 0.180 & 49 MPS$^\ddagger$ \\
SweRank$^\dagger$ & 43.0 [37.5, 48.7] & 100.0 & 97.3 & 44.2 & 13,339 & 9,370 & 96.3 & 0.095 & 31 A100 \\
{\color{gray}\textit{Gold files (upper bound)}} & {\color{gray}\textit{49.3 [43.7, 55.0]}} & {\color{gray}\textit{100.0}} & {\color{gray}\textit{97.7}} & {\color{gray}\textit{50.5}} & {\color{gray}\textit{12,206}} & {\color{gray}\textit{6,556}} & {\color{gray}\textit{not fitted}} & {\color{gray}\textit{0.075}} & {\color{gray}\textit{0}} \\
{\color{gray}\textit{No source (lower bound)}} & {\color{gray}\textit{8.7 [6.0, 12.4]}} & {\color{gray}\textit{100.0}} & {\color{gray}\textit{42.0}} & {\color{gray}\textit{20.6}} & {\color{gray}\textit{0}} & {\color{gray}\textit{0}} & {\color{gray}\textit{not fitted}} & {\color{gray}\textit{0.071}} & {\color{gray}\textit{0}} \\
\bottomrule
\end{tabular}
\caption{Per-context details, GPT-5.6, 300 issues (file-level BM25 is in Table~\ref{tab:main}). Res.$|$app.: resolution among issues whose patch applied. In $\pm$2\%: share of issues whose context is within 2\% of $B_i$; ``not fitted'': contexts not cut to $B_i$. \$/issue: API cost of the whole pipeline for that context (localization draws, planner and patch call; list prices); SWE-Pruner's includes the localization that builds its input, SweRank's is the patch call alone. Build s: median wall-clock seconds per issue to build the context with a model-based compressor or retriever (A6000 GPU, or CPU for LLMLingua-2; SweRank on an A100). $^\ddagger$SWE-Pruner's GPT-5.6 contexts were built on an Apple-silicon laptop; on an A100 it takes 5 s. $^\dagger$: added after the main results. Bounds in grey italic.}
\label{tab:arms}
\end{table*}
\rev{Table~\ref{tab:arms} lists every GPT-5.6 context: resolution with its interval, how often the patch parses and applies, resolution among applied patches, context size, the share of issues within the budget window, cost and build time. It backs the applicability numbers of \S\ref{sec:lose}.}

\begin{table}[t]
\centering
\footnotesize
\setlength{\tabcolsep}{2.4pt}
\begin{tabular}{lrrrrr}
\toprule
 & \multicolumn{2}{c}{Resolved} & \multicolumn{3}{c}{Claude Opus 5} \\ \cmidrule(lr){2-3}\cmidrule(lr){4-6}
Context & GPT & Claude & App. & Res.$|$app. & Tok. \\
\midrule
Ours (CQ2-CARE) & 43.7 & 46.7 & 83.3 & 56.0 & 27.1K \\
Single draw (Q50) & 46.3 & 43.0 & 78.7 & 54.7 & 37.0K \\
Size-matched draw & 42.7 & 48.3 & 83.7 & 57.8 & 28.1K \\
LLMLingua-2 & 30.0 & 40.3 & 80.7 & 50.0 & 26.5K \\
LongLLMLingua & 35.0 & 29.0 & 63.0 & 46.0 & 26.7K \\
LongCodeZip & 37.3 & 36.0 & 73.3 & 49.1 & 26.2K \\
BM25 (chunks) & 29.0 & 34.0 & 82.3 & 41.3 & 26.9K \\
Dense (chunks) & 22.7 & 33.7 & 86.7 & 38.8 & 27.0K \\
BM25 (files) & 34.3 & 34.0 & 80.7 & n/a & 26.3K \\
SWE-Pruner$^\dagger$ & 41.7 & 39.3 & 73.7 & 53.4 & 27.0K \\
SweRank$^\dagger$ & 43.0 & 45.0 & 89.3 & 50.4 & 26.5K \\
{\color{gray}\textit{Gold files (upper)}} & {\color{gray}\textit{49.3}} & {\color{gray}\textit{46.3}} & {\color{gray}\textit{83.0}} & {\color{gray}\textit{55.8}} & {\color{gray}\textit{12.2K}} \\
{\color{gray}\textit{No source (lower)}} & {\color{gray}\textit{8.7}} & {\color{gray}\textit{13.0}} & {\color{gray}\textit{81.0}} & {\color{gray}\textit{16.0}} & {\color{gray}\textit{0.0K}} \\
\bottomrule
\end{tabular}
\caption{The 300 issues with GPT-5.6 and with Claude Opus 5 for every call (resolved \%). Applies, Res.$|$app.\ and Tokens are for Claude. $^\dagger$: added after the main results. Bounds in grey italic.}
\label{tab:claude}
\end{table}
\rev{Table~\ref{tab:claude} is the Claude Opus 5 counterpart of Table~\ref{tab:main}, with resolution for both models side by side. Appendix~\ref{app:claude} discusses it.}

\begin{table*}[t]
\centering
\footnotesize
\setlength{\tabcolsep}{3pt}
\begin{tabular}{lrrrrrrrrr}
\toprule
\multicolumn{10}{l}{\emph{(a) Resolved (\%) by issue creation date, terciles of 100 issues}} \\
 & & & \multicolumn{7}{c}{Context minus ours} \\ \cmidrule(lr){4-10}
Created & \begin{tabular}[b]{@{}c@{}}No\\source\end{tabular} & Ours & \begin{tabular}[b]{@{}c@{}}LLM-\\Lingua-2\end{tabular} & \begin{tabular}[b]{@{}c@{}}LongLLM-\\Lingua\end{tabular} & \begin{tabular}[b]{@{}c@{}}Long-\\CodeZip\end{tabular} & \begin{tabular}[b]{@{}c@{}}SWE-\\Pruner$^\dagger$\end{tabular} & \begin{tabular}[b]{@{}c@{}}BM25\\(chunks)\end{tabular} & \begin{tabular}[b]{@{}c@{}}Dense\\(chunks)\end{tabular} & \begin{tabular}[b]{@{}c@{}}Swe-\\Rank$^\dagger$\end{tabular} \\
\midrule
2025-01--2025-04 & 10 & 48 & $-$14 & $-$6 & $-$6 & $-$1 & $-$12 & $-$24 & $-$1 \\
2025-04--2025-10 & 8 & 41 & $-$14 & $-$6 & $-$3 & $-$3 & $-$17 & $-$15 & +3 \\
2025-10--2026-05 & 8 & 42 & $-$13 & $-$14 & $-$10 & $-$2 & $-$15 & $-$24 & $-$4 \\
\midrule newest $-$ oldest & & & +1 & $-$8 & $-$4 & $-$1 & $-$3 & 0 & $-$3 \\
\quad 95\% CI & & & [$-$10, +12] & [$-$18, +2] & [$-$15, +7] & [$-$10, +8] & [$-$15, +9] & [$-$13, +13] & [$-$14, +8] \\
\midrule
\multicolumn{10}{l}{\emph{(b) Within-issue regression of resolution, coefficient [95\% CI]; $^\dagger$ rows add SWE-Pruner and SweRank}} \\
 & \multicolumn{4}{r}{without applies} & \multicolumn{5}{r}{with applies} \\
\midrule
\multicolumn{3}{l}{Verbatim lines $\varepsilon$ (3,298 pairs)} & \multicolumn{3}{r}{+0.146 [+0.104, +0.188]} & \multicolumn{4}{r}{+0.043 [+0.005, +0.082]} \\
\multicolumn{3}{l}{\quad 3,898 pairs$^\dagger$} & \multicolumn{3}{r}{+0.151 [+0.109, +0.194]} & \multicolumn{4}{r}{+0.050 [+0.012, +0.088]} \\
\multicolumn{3}{l}{Gold-file recall $\rho$ (3,298 pairs)} & \multicolumn{3}{r}{+0.273 [+0.213, +0.333]} & \multicolumn{4}{r}{+0.213 [+0.157, +0.269]} \\
\multicolumn{3}{l}{\quad 3,898 pairs$^\dagger$} & \multicolumn{3}{r}{+0.260 [+0.200, +0.320]} & \multicolumn{4}{r}{+0.206 [+0.150, +0.261]} \\
\multicolumn{3}{l}{log tokens (3,298 pairs)} & \multicolumn{3}{r}{$-$0.030 [$-$0.053, $-$0.006]} & \multicolumn{4}{r}{$-$0.031 [$-$0.053, $-$0.008]} \\
\multicolumn{3}{l}{\quad 3,898 pairs$^\dagger$} & \multicolumn{3}{r}{$-$0.032 [$-$0.055, $-$0.008]} & \multicolumn{4}{r}{$-$0.033 [$-$0.055, $-$0.010]} \\
\multicolumn{3}{l}{Patch applies (3,298 pairs)} & \multicolumn{3}{r}{not in model} & \multicolumn{4}{r}{+0.342 [+0.277, +0.407]} \\
\multicolumn{3}{l}{\quad 3,898 pairs$^\dagger$} & \multicolumn{3}{r}{not in model} & \multicolumn{4}{r}{+0.333 [+0.270, +0.397]} \\
\bottomrule
\end{tabular}
\caption{\textbf{Post-hoc offline checks} (GPT-5.6, 300 issues). (a) Resolution (\%) of the no-source context and of ours by issue creation date, and each baseline's resolution minus ours (points); a gap driven by memorized fixes would change on newer issues. (b) Linear probability model of resolution with issue fixed effects over every (issue, context) pair that shows source, without and with the code-specific baselines; the right column adds whether the patch applies. Intervals: 20,000-draw issue bootstrap. $^\dagger$: added after the main results.}
\label{tab:w2}
\end{table*}
\rev{Table~\ref{tab:w2} holds the two offline checks: the split by issue creation date used in the Limitations, and the within-issue regression of \S\ref{sec:controlled}, each with and without the two trained methods.}

\end{document}